\documentclass[twocolumn,prb,amsmath,amssymb]{revtex4}
\usepackage{graphicx}
\usepackage{dcolumn}
\usepackage{bm}
\usepackage{psfrag} 
\usepackage{mathptmx}
\usepackage{mathrsfs}

\def\half{\frac{1}{2}}
\def\tF{{\mbox{\tiny{F}}}}
\def\tB{{\mbox{\tiny{B}}}}
\def\bb{\mathbf}
\def\bk{\mathbf{k}}
\def\vr{\mathbf{r}}

\def\state#1{|#1 \text{ Fermions} \rangle}
\def\bra#1{\langle{#1} |}
\def\ket#1{|{#1} \rangle}
\def\PLLL{\mathcal{P}_{\mbox{\tiny{LLL}}}}

\def\etal{{\it et al.~}}

\def\beq{\begin{equation}}
\def\eeq{\end{equation}}

\def\ud{{\uparrow\downarrow}}
\def\dd{{\downarrow\downarrow}}
\def\uu{{\uparrow\uparrow}}
\def\trialIdx{\gamma}  
\def\corrSym{h}
\newcommand\derivX[1]{%
\frac{\partial }{\partial {#1}}}

\begin{document}

\title{Trial Wavefunctions for $\nu= \frac{1}{2} + \frac{1}{2}$ Quantum Hall Bilayers}

\author{Gunnar M\"oller,${}^{1}$ Steven H. Simon${}^2$ and Edward H. Rezayi${}^{3}$}
\affiliation{ ${}^1$ TCM Group, Cavendish Laboratory, J.~J.~Thomson Ave., Cambridge CB3 0HE, UK \\
${}^2$ Bell Laboratories, Alcatel-Lucent,  Murray Hill, New Jersey 07974\\
${}^3$ Department of Physics, California State University,
 Los Angeles, California 90032
}
\date{November 25th, 2008}

\begin{abstract}
Quantum Hall bilayer systems at filling fractions near
$\nu=\frac{1}{2} + \frac{1}{2}$ undergo a transition from a compressible phase
with strong intralayer correlation to an incompressible phase with
strong interlayer correlations as the  layer separation $d$ is
reduced below some critical value.   Deep in the intralayer phase
(large separation) the system can be interpreted as a fluid of
composite fermions (CFs), whereas deep in the interlayer phase
(small separation) the system can be interpreted as a fluid of
composite bosons (CBs).  The focus of this paper is to understand
the states that occur for intermediate layer separation by using
trial variational wavefunctions.  We consider two main classes of
wavefunctions.  In the first class,
previously introduced in M\"oller \etal [Phys.~Rev.~Lett.~{\bf 101}, 176803 (2008)],
we consider interlayer BCS pairing of two
independent CF liquids.  We find that these wavefunctions are
exceedingly good for $d \gtrsim \ell_0$ with $\ell_0$ the magnetic
length.   The second class of wavefunctions naturally follows the
reasoning of Simon \etal [Phys.~Rev.~Lett.~ {\bf 91}, 046803 (2003)] and generalizes the
idea of pairing wavefunctions by allowing the CFs also to be
replaced continuously by CBs.  This generalization allows us to
construct exceedingly good wavefunctions for interlayer spacings of
$d \lesssim \ell_0$, as well. The accuracy of the wavefunctions
discussed in this work, compared with exact diagonalization, 
approaches that of the celebrated Laughlin wavefunction.  
\end{abstract}
\maketitle

\section{Introduction}
\label{sec:BilayerIntro}

In bilayer quantum Hall systems at filling fraction $\nu=\half +
\half$,  at least two different quantum states of matter are known
to occur, depending upon the spacing $d$ between the
layers.\cite{DasSarmaPerspectives} For large enough spacing, the two
layers interact very weakly and must be essentially independent
$\nu=\half$ states, which can be described as compressible composite
fermion (CF) Fermi seas.\cite{Heinonen}  So long as the distance
between the two layers is very large, there are very strong
intralayer correlations but very weak interlayer correlations
(although, as we will discuss below, even very weak interlayer
correlations may create a pairing instability at exponentially low
temperatures \cite{BonesteelPRL96}). Conversely, for small enough
spacing between the two layers the ground state is known to be the
interlayer coherent ``111 state'', which we can think of as a
composite boson (CB), or interlayer exciton
condensate,\cite{DLReviewNature} with strong interlayer correlations
and intralayer correlations which are weaker than those of the
composite fermion Fermi sea.\cite{DasSarmaPerspectives} While the
nature of these two limiting states is reasonably well understood,
the nature of the states at intermediate $d$ is less understood and
has been an active topic of both
theoretical\cite{JoglekarPRB01,SchliemannPRL01,SchliemannPRB03,SternandHalperinPRL02,
ParkPRB04,YoshiokaPRB66,KimNayak01,BonesteelPRL96,MorinariPRB99,USPRL,JinwuYePRL,
doretto06,Milovanovic07} and experimental
interest.\cite{MurphyPRL94,PRL84Spielmanetal00,PRL87Spielmanetal01,
PRL88Kelloggetal02,PRL90Kelloggetal03,Spielmanetal04,PRL03Tutuc04,
Wiersmaetal04,Kumadaetal05,champagne08PRL,Eisenstein08PRB}
Although there are many interesting questions remaining that involve
more complicated experimental situations, within the current work we
always consider a zero temperature bilayer system with zero
tunnelling between the two layers and no disorder. Furthermore, we
only consider the situation of $\nu = \half + \half$ where the
electron density in each layer is such that $n_1 = n_2 = B /(2
\phi_0)$ with $\phi_0 = h c/e$ the flux quantum and $B$ the magnetic
field. Finally we assume that electrons are fully spin-polarized, we
neglect the finite extension of the wave functions in the
$z$-direction, and we always assume that the magnetic field is
precisely perpendicular to the plane of the sample.

Our main focus in this work is on the nature of the transition
between interlayer 111 (CB) state and the intralayer Fermi liquid
(CF) state.  Currently, contradictory conclusions about
the nature of the transition may be drawn from the literature.  The
experiments are complex and are frequently hard to interpret (and
may require assumptions beyond the simplifying assumptions made in
the current paper). While some of the experiments\cite{MurphyPRL94,
PRL84Spielmanetal00,PRL87Spielmanetal01,
PRL88Kelloggetal02,PRL90Kelloggetal03,Spielmanetal04}
point towards a continuous transition between two phases, it is not
clear whether this could actually be a first order transition
smeared by disorder.\cite{SternandHalperinPRL02}  There is no
doubt, however, that a notable change of behavior takes place
in the approximate vicinity of $d/\ell_0 \approx 1.7$ with
$\ell_0 = \sqrt{\phi_0/B}$ as the magnetic length.

Theoretically, the situation has also remained unclear. Several theoretical works\cite{SchliemannPRL01,Joglekar04,MoonetalPRB95} found indications
of a first order transition near $d/\ell_0 \approx 1.3$, whereas
others have found no indication for a first order transition and
evoke a continuous evolution of correlations,\cite{YoshiokaPRB66}
and indications of a continuous transition
occuring near $d/\ell_0 \approx 1.6$.\cite{Yoshioka06}

Description of the phases that occur in the bilayer system has also been quite a challenge. Some very
influential works have pointed to the possibility that a
number of exotic phases could be lurking within this
transition as well.\cite{KimNayak01,ParkPRB04,BonesteelPRL96,BonesteelPRB93,YoshiokaPRB66,JinwuYePRL}
In particular, it has been suggested\cite{BonesteelPRL96,KimNayak01,MorinariPRB99} that the bilayer CF
Fermi sea is always unstable to BCS pairing from weak interactions
between the two layers (due to gauge field fluctuations).   Some of
these works\cite{KimNayak01,MorinariPRB99} further concluded that
the pairing of CFs should be in the $p_x - i p_y$ channel, which would be
analogous to the pairing that occurs in single layer CF systems to
form the Moore-Read Pfaffian state\cite{MooreRead91,Greiter92}
from the CF Fermi sea.     However, these works did not provide any numerical evidence supporting these claims.

Recent work by the current authors\cite{USPRL} has shed considerable light on the subject. In this work, compelling numerical evidence was given that for $d/\ell_0 \gtrsim 1$ the ground state is well described as a CF-BCS paired phase, although the pairing channel is $p_x + i p_y$ rather than the previously predicted $p_x - i p_y$.  Explicit pairing wavefunctions were shown to have extremely high overlaps with the exact ground state for small systems.
This work will be described in more detail below.

A somewhat different approach has also been proposed by
some of the present authors and collaborators\cite{PRLSimon03} in order to understand the transition between the phase at large $d$ and the 111, or CB phase at small $d$.
In that work, a set of trial wavefunctions was
constructed to attempt to describe the crossover.
This theory (to be discussed in depth below)
provides an intuitive picture for the transition from the CF-liquid
product state to the 111-state in terms of an energy trade-off
between intralayer interaction energy and interlayer interaction
energy. At large layer separation $d$, CFs fill a Fermi sea. These
CFs can be thought of as electrons bound to a pair of correlation holes
within the same layer. At small layer separation the 111-state can
be thought of as a condensate of interlayer excitons or composite
bosons. These composite bosons are formed by electrons bound to
a correlation hole in the opposite layer, which is
in fact a true hole of charge $+e$, just as a Laughlin quasihole on
top of a $\nu=1$ quantum Hall liquid. Additionally, CBs carry a single
correlation hole in the same layer. Within the theory of
Ref.~\onlinecite{PRLSimon03}, at intermediate $d$ wavefunctions were
introduced with some density of CFs having particle-hole binding within
the layer and some density of CBs having particle-hole binding between
layers.  As the distance $d$ between the layers is continually
reduced, the CFs are continually replaced by CBs and the intralayer
correlation is replaced by interlayer correlations.

While physically appealing, this description of the transition is
clearly incomplete in that it considers CFs and CBs as independent
types of particles, though in reality all of the electrons must be
identical.   Both the CFs and CBs consist of electrons bound to
correlation holes or vortices, or with  ``flux attached" in the
Chern-Simons language. The difference between the CBs and CFs is
whether they are bound to correlation holes in the opposite
layer (CBs) or only within the same layer (CFs). However, nothing
prevents electrons from breaking free from their correlation holes
and becoming bound to a different correlation hole --- which could
then change the identity of a particle from a CB to a CF and vice
versa. Indeed, whenever two composite bosons in opposite layers
approach the same coordinate position, they can ``trade'' their
accompanying correlation holes (vortices or flux quanta), and emerge
as two composite fermions. In terms of a second quantized notation,
with $\psi$ representing a composite fermion annihilation operator,
and $\phi$ representing a composite boson annihilation operator,
such scattering processes would be described by an interaction term
\begin{equation}
\lambda_{\mathbf{k}_1, \mathbf{k}_2, \mathbf{k}_3, \mathbf{k}_4} \,\,
\psi^\dagger_{\uparrow,\mathbf{k}_1}\psi^\dagger_{\downarrow,\mathbf{k}_2}
\phi_{\uparrow,\mathbf{k}_3}\phi_{\downarrow,\mathbf{k}_4} \;+\;
h.c.
\end{equation}
with $\uparrow$ and $\downarrow$ representing the two layers and
$\lambda$ as a coupling constant (and $h.c.$ denoting the hermitian
conjugate). If the bosons happen to be condensed, there is a
large expectation for the CBs to be in a $\mathbf{k}=0$ state.
Invoking momentum conservation,  the most dominant such interaction
term is then of the form
\begin{equation}
\lambda_k  \,\,
\psi^\dagger_{\uparrow,\mathbf{k}_1}\psi^\dagger_{\downarrow,\mathbf{-k}_1}
\langle \phi_{\uparrow,\mathbf{k=0}}\phi_{\downarrow,\mathbf{k=0}}
\rangle   \; + \;h.c.
\end{equation}
which we immediately recognize as a pairing term for the composite
fermions.   Thus we see that the mixed CF-CB picture is quite closely linked to the idea of CFs forming a CF-BCS state.

As discussed above, our numerics indicate that CF pairing occurs in the $p_x+ip_y$ channel.  An equivalent statement is that the two--CF pair wavefunction
acquires a phase of $+2 \pi$ as two paired electrons in opposite
layers are taken in a clockwise path around each other.    We will further argue that this is the only pairing symmetry that is compatible with coexistence of CFs and CBs.
The argument rests on the fact that for the 111 wavefunction, taking any electron
around any other electron in the opposite layer will result in a $+2
\pi$ phase. As will be further illustrated below, compatibility of CBs
that make up the 111 state with the CFs that compose the $p$-wave
paired CF state requires that these phases match, and will require that the
$p$-wave pairing is of $p_x+ip_y$ type.

In the current work, we construct explicit wavefunctions for
interlayer paired CF states.  As in BCS theory, the shape of the
pairing wavefunction is treated in terms of a set of (a very small
number of) variational parameters.  As previously discussed in Ref.~\onlinecite{USPRL} we find that for interlayer spacings $d \gtrsim \ell_0$ our trial states are exceedingly good
representations of the ground state.   However, at spacings below $d
\approx \ell_0$ we find that the simple paired CF states are no
longer accurate.  We then return to the above described idea of
CF-CB mixtures.   With only one additional variational
parameter representing the probability that an electron is a CB
versus being a CF, we obtain a family of wavefunctions that nearly
match the exact ground state for all values of $d/\ell_0$.

The general structure of this paper is as follows.  In section
\ref{sec:Theory}, we will discuss in detail the particular
wavefunctions to be studied.  First, in section \ref{sec:CF_liquid}
we review the composite fermion Fermi liquid in single layer
systems, and focus on some particular aspects that help us construct
bilayer states with paired CFs, previously introduced in Ref.~\onlinecite{USPRL}, in section \ref{sec:TheoryPairedCF}.  We then turn to the discussion of the interlayer coherent 111-state in section \ref{sec:111-state} and how it too can be interpreted as
both a state of composite bosons (CBs) and as a paired state. In
section \ref{sec:MixedCFCBTheory} we discuss the merging of the
physics of CBs with that of the paired CF states to yield a mixed
fluid wavefunction which incorporates both types of physics.
Crucially, we show in this section that $p_x + i p_y$ is the only
pairing symmetry of CFs that can coexist with CBs.  We note that
wavefunctions discussed in section \ref{sec:MixedCFCBTheory} include
the mixed CB-CF wavefunctions of Ref.~\onlinecite{PRLSimon03} as a
special case.

Having constructed a family of variational wavefunctions, we proceed
to test the validity of this approach based on numerical
calculations on the sphere presented in section
\ref{sec:numericalResults}. Data from Monte-Carlo simulations of the
paired CF and mixed fluid wavefunctions is compared with data
obtained from exact numerical diagonalizations of the Coulomb
Hamiltonian for model systems of up to 14 electrons in sections
\ref{sec:pairedResults} and \ref{sec:mixedResults}. In section
\ref{sec:occupation}, we discuss the properties of the various trial
states via the occupation of CF orbitals and in analogy to a
BCS superconductor. Section \ref{sec:symmetries} is devoted to a
discussion of order parameters that characterize the system.  In
Section \ref{sec:discussion} we further discuss our undestanding and
interpretation of our results.  We also briefly discuss a number of
issues including the effects of finite temperature, layer density
imbalance, tunneling between the layers, and electron spin. 
Consequences for electronic transport are also analyzed. Finally
we discuss the expected transport properties of the phases we
describe.  In section \ref{sec:conclusion} we conclude and briefly
summarize our results.

We have relegated to the appendices a number of details that are not in
the main development of the paper.   In Appendix
\ref{sec:CFCBOnSphere}, we discuss in detail how to adapt the mixed
fluid wavefunctions to obtain a representation on the sphere. More
numerical results for a restricted class of wave functions,
corresponding to filled CF shells on the sphere, are discussed in
Appendix \ref{sec:AppFilledShells}. Further details about the procedure
applied for the optimization of trial states are elaborated in
Appendix \ref{sec:numericalmethods}. Finally, Appendix \ref{sec:correlations}
discusses some properties of the two-electron correlation functions
in the bilayer system.

\section{Wavefunctions for the quantum Hall bilayer}
\label{sec:Theory} In this section we review the various trial
wavefunctions that we will be studying throughout this paper.
To the experienced reader the discussion of the composite
fermion liquid (section \ref{sec:CF_liquid}) and the 111 state
(section \ref{sec:111-state}) will be mostly review. This material
is nonetheless included in depth to emphasize a few key points that
guide our reasoning.

For simplicity, in this section we will consider infinite-sized
systems on a planar geometry so that we can write wavefunctions in
the usual complex coordinate notation.  Here and in the following,
$z_i = x_i +i y_i$ is the complex representation of the
coordinates of particle $i$ (with the overbar representing the complex
conjugate), and the usual Gaussian factors of $e^{-\sum_i z_i \bar
z_i /(2\ell_0)^2}$ are understood to be included in the measure of
the Hilbert space and will not be written explicitly for simplicity
of notation. For bilayer states, we note coordinates in the second
layer as $w_j$, using the same complex representation. In section
\ref{sec:numericalResults} below, we will convert to considering
wavefunctions on the sphere, where we actually perform our numerical
calculations. The changes required to adapt our theory to the spherical
geometry are discussed in Appendix \ref{sec:CFCBOnSphere}.

\subsection{Composite Fermion Liquid}
\label{sec:CF_liquid}

For bilayer systems at infinite layer spacing, the interlayer
interaction vanishes and the two layers can be considered as
independent $\nu=\half$ systems.   For such single layer $\nu=\half$
systems, the composite fermion approach\cite{Heinonen} has been
remarkably successful in describing a great deal of the observed
physics.   In this picture,\cite{EarlyJain,Heinonen} the wave
function for interacting electrons in magnetic field $B$ is written
in terms of the wavefunction for free (composite) fermions in an
effective magnetic field ${\cal B} = B - 2 n \phi_0$ with the
density of electrons $n$.   Each fermion is also attached to two
vortices (or correlation holes) of the wavefunction (Jastrow
factors) resulting in the following type of wavefunction:
\begin{equation}
\label{eq:simpleCFs} \Psi^{CF}=\PLLL \prod_{k<p}(z_p-z_k)^2
\det \left[\phi_i(z_j,\bar z_j)\right].
\end{equation}
where $\phi_i$ are the orbitals for free fermions in the effective
magnetic field $\cal B$, and $\PLLL$ is the projection operator
that projects to the lowest Landau level. The determinant in
equation (\ref{eq:simpleCFs}) above describes a Slater determinant
of electrons at $z_i$ filling states given by the orbitals $\phi_i$.

    For the special case $\nu=\frac{1}{2}$ the CFs experience zero
effective field and behave similarly as electrons at zero field,
forming a Fermi sea.\cite{HLR,Willett93,Heinonen} For an infinitely
extended plane, plane waves form a basis of single particle orbitals
for particles in zero effective magnetic field such that
\begin{equation}
\phi_i(z_j)=e^{i\bb k_i \cdot \bb r_j}.
\end{equation}
Since $\bb k \cdot \bb r = \frac{1}{2}(k\bar z + \bar k z)$ (with
$k$ being the complex representation of the vector $\bb k$)  and the
projection on the LLL transforms $\bar z \rightarrow
-2\frac{\partial}{\partial z}$, the plane wave factors become
translation operators under projection.\cite{Read94} This yields
\begin{equation}
\Psi_{\frac{1}{2}}=\mathcal{A}\left\{ \prod_{i<j} \bigl(
[z_i+\ell_0^2k_i] - [z_j+\ell_0^2k_j]\bigr)^2  \prod_i e^{i\bar
k_i z_i/2}\right\},
\end{equation}
where $\mathcal A$ is the antisymmetrizing operator that sums over
all possible pairings of the $z_i$'s with the $k_j$'s, odd
permutations added with a minus sign.   We see that the fermions are
still bound to zeros of the wavefunction, but the positions of the
zeros (correlation holes) are moved away from the electrons by a
distance $\ell_0^2 k$, which is given in terms of ``momentum'' $k$.
In order to minimize the Coulomb energy, these distances should be
minimized, but simultaneously, all the $k_i$ have to be different or
the wavefunction will vanish on antisymmetrization. Thus, to
minimize potential energy, the $k_i$'s fill up a Fermi sea of
minimal size. This is how the potential energy becomes the driving
force for establishing the Fermi sea. Although this naive picture of
charged dipole dynamics is not strictly true in the way that it is
presented here,\cite{JainDipole} there are several ways to more
rigorously embody this type of dipolar Fermi sea dynamics in a
theory of the lowest Landau level, which give credibility to this
type of simplified argument.\cite{ShankarReview,ReadDipole,SternDipole}

Unfortunately, the projection ${\cal P}$ in Eq.\ \ref{eq:simpleCFs}
is exceedingly difficult to implement numerically for large systems.
To circumvent this problem, Jain and Kamilla\cite{PRBJainKamilla97} proposed a
rewriting of the composite fermion wavefunction as
\begin{equation}
\label{eq:Cf2}
    \Psi^{CF} = \prod_{k<p}(z_p-z_k)^2\, \det \left[\tilde \phi_i(z_j) \right],
\end{equation}
where
\begin{equation}
\label{eq:tildephi}
    \tilde \phi_i(z_j) = J_j^{-1}\PLLL \left[\phi_i(z_j,\bar z_j) J_j \right],
\end{equation}
with $J_j=\prod_{k\neq j}(z_k-z_j)$ and the $\phi_i$ chosen such as to
represent wavefunctions corresponding to a filled Fermi sea.\cite{TildeNotation} This
form, while not strictly identical to the form of Eq.~\ref{eq:simpleCFs},
is extremely close numerically and has equally
impressive overlaps with exact diagonalizations\cite{PRBJainKamilla97} and is therefore an equally good starting
point for studying composite fermion physics. However, in contrast
to the form of Eq.\ \ref{eq:simpleCFs}, the form of Eqs.\ \ref{eq:Cf2}
and \ref{eq:tildephi} are comparatively easy to evaluate numerically
and therefore allow large system quantum Monte-Carlo calculations.\cite{PRBJainKamilla97}
In this paper, we have used this type of approach.

In order to obtain a wavefunction for the bilayer system
at $\nu=\half+\half$ and infinite layer separation, a simple product state of
two composite fermion liquids (CFL) is appropriate.
\beq
\label{eq:ProductState}
   \Psi(d\to\infty)=|CFL\rangle \otimes |CFL\rangle
\eeq At finite layer separation, however, correlations between the
layers are expected to exist and have been suspected to resemble a
paired state.\cite{BonesteelPRB93,BonesteelPRL96,MorinariPRL98,MorinariPRB99,KimNayak01}
As we will see below, the product state (\ref{eq:ProductState}) may
be regarded as a particular paired state whenever the Fermi-surface
is inversion symmetric with respect to $\mathbf{k}=0$, i.e., the center 
of the Fermi-sea. 
In these cases, for each particle in layer one occupying a
state with momentum $\bk$, there exists its partner in layer two occupying
a state with momentum $-\bk$.

\subsection{Paired CF bilayer state}
\label{sec:TheoryPairedCF}

We now consider how to write a trial wavefunction for an interlayer
paired composite fermion state, which we suggest should be an
accurate description of the bilayer system when the spacing between
the layers is large.   The material in this section is mostly a review 
of material introduced in Ref.~\onlinecite{USPRL}.
 As a starting point, let us take the well
known BCS wavefunction in the grand canonical ensemble\cite{deGennesSupra} \beq \label{eq:BCS} \ket{\Psi} = \prod_\bk
\left( u_\bk + v_\bk \, e^{i\varphi} \, a^\dagger_{\bk\downarrow}
a^\dagger_{-\bk\uparrow}\right)\ket{0} \eeq with the normalization
$|u_\bk|^2 + |v_\bk|^2 =1$ and where $a^\dagger_{\bk\uparrow}$
creates a particle in layer $\uparrow$ with momentum $\bk$.   Note
that the $u$'s and $v$'s are properly understood here as variational
parameters of the BCS wavefunction. Next, we rewrite this
wavefunction in an unnormalized form by multiplying all factors by
$u_\bk^{-1}$ and defining $g_\bk=v_\bk/u_\bk$, so \beq
\label{eq:BCS1a} \ket{\Psi} = \prod_\bk \left( 1 + g_\bk \,
e^{i\varphi} \, a^\dagger_{\bk\downarrow}
a^\dagger_{-\bk\uparrow}\right)\ket{0}. \eeq

Finally, we project to a
fixed number $2N$ of particles (i.e, switch to canonical ensemble)
by integration over $\int d\varphi\,\exp(-iN\varphi)$ such that we
retain exactly $N$ pair creation operators. This yields \beq
\label{eq:BCS2} \ket{\Psi} = \sum_{\{\bk_1,\ldots,\bk_N\}}
\prod_{\bk_i} \, g_\bk \, a^\dagger_{\bk_i\downarrow}
a^\dagger_{-\bk_i\uparrow}\ket{0}. \eeq In the first quantized
language, we can write
\begin{subequations}
\beq \label{eq:BCS_RealSpace} \Psi =
 \det \left[
g(\vr_{i\downarrow},\vr_{j\uparrow}) \right]  \eeq 
where $g$ is the Fourier transform \beq
g(\vr_{i\downarrow},\vr_{j\uparrow}) = \sum_\bk g_\bk \,
e^{i\bk\cdot(\vr_{i\downarrow}-\vr_{j\uparrow})}. \eeq
\end{subequations}
Note that the exponential factor of the Fourier transform can be
regarded as a product of two basis functions $\phi_\bk(\vr) = e^{i
\bk \vr}$ on the plane, i.e.\
\begin{equation}
e^{i\bk\cdot(\vr_{i\downarrow}-\vr_{j\uparrow})} = e^{i\bk\cdot
\vr_{i\downarrow}} e^{-i \bk\cdot \vr_{j\uparrow}} =
\phi_\bk(\vr_{i\downarrow}) \phi_{-\bk}(\vr_{j\uparrow}).
\end{equation}
With this in mind, similar paired wavefunctions can be written
for more general geometries with arbitrary basis functions. In the
following, we construct paired states for composite fermions in
the bilayer system (denoting particles in the upper layer as $z$ and
those in the lower layer as $w$). As in section \ref{sec:CF_liquid} we will multiply our
fermion wavefunction with composite-fermionizing Jastrow factors
and project to the lowest Landau level yielding
\begin{align}
\label{eq:BCS_RealSpaceCF}
\Psi &= \PLLL \prod_{i<j}(z_i-z_j)^2 \prod_{i<j}(w_i-w_j)^2 \det\left[g(z_i,w_j)\right] \nonumber\\
&\equiv \PLLL \det \left[J^{zz}_i J^{ww}_j g(z_i,w_j)\right],
\end{align}
where we have defined ``single particle" Jastrow factors
\begin{subequations}
\begin{eqnarray} J^{zz}_i &=& \prod_{k\neq i} (z_i-z_k) \\J^{ww}_i &=& \prod_{k\neq i} (w_i-w_k).
 \end{eqnarray}
\end{subequations}
In order to handle the projection numerically, we follow the recipe
of Jain and Kamilla (\ref{eq:tildephi}) discussed above, bringing
the Jastrow factors inside the determinant and projecting individual
matrix entries. This prescription applies to the bilayer case in a
similar manner as for the single layer case  (since the total
Hilbert space of the bilayer system may be represented as a direct
product of the space for each layer and projection in one space does
not affect the other).    We then obtain the final paired wave
function: \begin{subequations}\label{eq:AllBilayerPaired} \beq
\label{eq:BilayerPaired} \Psi^\text{CF-BCS} = \det \left[ g_\tF(z_i,
w_j) \right] \eeq where
\begin{equation}
\label{eq:gsum}  g_\tF(z_i, w_j) = \sum_\bk g_\bk \, J_i^{zz}\,
J_j^{ww}\, \tilde\phi_\bk(z_i) \, \tilde\phi_{-\bk}(w_j).
 \end{equation}
 \end{subequations}
We denote the projected CF orbitals $\tilde\phi$ as defined
in equation (\ref{eq:tildephi}) above.
By convention, the single particle Jastrow factors $J_i$ are kept
inside the function $g_\tF$ so that $g_\tF(z_i - w_j)$ is actually a
function of all of the $z$'s and $w$'s through the $J$'s.\cite{TildeNotation}  
The subscript \mbox{\footnotesize{F}} here has been chosen to indicate 
that these are paired composite {\bf F}ermions.   Note that in the above
expressions $\bk$ may stand for a general set of orbital quantum
numbers (this will be important for spherical geometry where the
free wavefunctions are spherical harmonics rather than plane
waves).

The $g_{\bk}$'s defining the shape of the pair wavefunction are
variational parameters, analogous to the usual $u$'s and $v$'s.
These parameters must be optimized to obtain a good wavefunction,
although the optimal solution will certainly depend on the layer
separation $d$. We also note that the expression
(\ref{eq:AllBilayerPaired}) can describe pairing in arbitrary
pairing channels depending upon the choice of $g_{\bk}$ and the
basis set $\{\phi_\bk\}$.   As a general definition, when the pair
wavefunction has the short distance form \beq
\label{eq:lWavePairing} g(z_i,w_j)\propto (z_i-w_j)^l \times h(
|z_i-w_j| ), \eeq with $h(0) \neq 0$, we say this is $l$-wave pairing. However, note
that $g(z,w)$ should asymptotically approach zero for
$|z-w|\to\infty$, such that the pair wavefunction can be normalized.
We also frequently use the atomic physics nomenclature where $l=0$
is termed $s$-wave, $l=1$ is the $p$-wave, and so forth.
Furthermore, $l=+1$ is denoted as $p_x + i p_y$ pairing, whereas
$l=-1$ is $p_x - i p_y$ pairing.   (Unfortunately, in the literature 
``$p_x + i p_y$" is used to denote either chirality). 
Note that the pairing symmetry is independent
of whether we move the $J^{zz}$ and $J^{ww}$ factors inside or
outside of the function $g_\tF$.

The choice of the pairing channel $l$ affects the precise value of
the flux $N_\phi=2(N-1)+l$ at which the trial state (\ref{eq:AllBilayerPaired})
occurs. For systems with finite $N$, we can thus distinguish the 
different possible pairing channels by studying the flux $N_\phi$ for which
the groundstate of the system is incompressible as a function of system size $N$.
Such a study has been undertaken in depth in Ref.~\onlinecite{USPRL} and it was 
found clearly that $p_x+ip_y$ pairing is supported by the numerical data. 
As the effective interaction of composite fermions
derives from the interaction of the underlying electrons in a non-trivial
manner, the pairing channel realized in the bilayer system was not
reliably predicted by various theoretical 
approaches.\cite{BonesteelPRB93,BonesteelPRL96,KimNayak01}

A case of particular interest is when the variational parameters
$g_\bk$ are defined as follows: \beq \label{eq:CFL_asPairedState}
g_\bk=\left\{
\begin{array}{ll}
\mbox{anything nonzero},& |\bk| \leq k_F \\
0,& \text{otherwise}
\end{array}
\right. \eeq It is easy to show that this choice of variational
parameters recovers the product state of two composite fermion
liquids (Eq.~\ref{eq:ProductState}).

\subsection{111-state}
\label{sec:111-state}

When the spacing between the two layers becomes small, the bilayer
system forms an interlayer coherent state.  A number of different
approaches have been used to understand this state and a
large amount of progress has been made using a mapping
to an iso-spin easy-plane ferromagnet.\cite{MoonetalPRB95,MoonetalPRB96}
In this work, however, we will
follow the Laughlin approach of considering trial wavefunctions in
a first quantized description. When the distance between the two
layers becomes zero, the exact ground state wavefunction of $\nu =
\half + \half$ is known to be the so-called 111-state\cite{Halperin111,MoonetalPRB95}
\begin{equation}
\label{eq:111state}
\Psi_{111} = \prod_{i<j}(z_i-z_j)\prod_{k<l}(w_k-w_l)\prod_{r,s}(z_r-w_s),
\end{equation}
where again we use $z$ to represent particles in the upper layer and
$w$ to represent particles in the lower layer. In contrast to the CF
state, (\ref{eq:111state}) contains only one Jastrow factor between
particles in the same layer so that the wavefunction is properly
antisymmetric under exchange of particles in the same layer.  Thus,
no additional determinant is needed to fix the symmetry as was the
case in the CF state.  In addition, (\ref{eq:111state}) includes a
Jastrow factor between particles in opposite layers. Consequently,
there is no amplitude for finding two particles at the same position
in opposite layers.   This can be interpreted as each particle being
bound to a hole in the neighboring layer.   One can say the
111-state is composed of interlayer excitons.\cite{DLReviewNature}
Another terminology is the Chern-Simons language where the electrons
are transformed into bosons bound to flux quanta, where each flux
quantum penetrates both layers. These ``composite bosons" can be
thought of as an electron bound to a vortex of the wavefunction in
each layer. Condensing these bosons gives the wavefunction
$\Psi_\text{CB}=1$ for the composite particles and the transform
back to an electron wavefunction (by reattaching the Jastrow
factors) yields (\ref{eq:111state}).

However, it is also useful to rewrite the
111 wavefunction using the Cauchy identity
\begin{equation}
\label{eq:CauchyId}
\prod_{i<j}(z_i-z_j) \prod_{i<j}(w_i-w_j)= \prod_{i,j}(z_i-w_j) \det\frac{1}{z_i-w_j}
\end{equation}
which yields
\begin{equation}
\label{eq:Paired111}
\Psi_{111} = \det\left[\frac{1}{z_i-w_j}\right]\prod_{i,j}(z_i-w_j)^2.
\end{equation}
This notation resembles the form of a general paired bilayer state as
discussed above in section \ref{sec:TheoryPairedCF}.  This
resemblance has been noted previously,\cite{KimNayak01,Ho95} and
from the form of the  $1/(z_i-w_j)$ factor, it has been concluded
that the pairing symmetry is $(p_x-ip_y)$.\cite{KimNayak01} Here, we
would like to propose a different interpretation.  Since the Jastrow
factors outside the determinant cancel the apparent singularity in
Eq.~\ref{eq:Paired111}, the phase obtained by taking an electron
around its partner is actually $+2 \pi$ rather than $-2 \pi$.  In
fact, for the 111 state it is clear from the explicit form
(\ref{eq:111state}) that as any electron is taken around another
electron in either layer, one accumulates a phase of precisely
$+2\pi$.  For clarity, it is useful to move the Jastrow factors in
Eq.~\ref{eq:Paired111} inside the determinant. We obtain

\begin{equation}
\label{eq:my_111paired} \Psi_{111}= \det \left[ g_\tB(z_i,w_j)
\right]
\end{equation}
where
\begin{equation}
g_\tB(z_i,w_j)  = \frac{J^{zw}_i J^{wz}_j}{z_i-w_j}
\end{equation}
and the interlayer partial Jastrow factors are defined by
\begin{subequations}
\begin{eqnarray}
\label{eq:partialJastrow1} J^{zw}_i &=& \prod_k (z_i-w_k) \\
             J^{wz}_i &=&\prod_k (w_i-z_k).
\end{eqnarray}
\end{subequations}
Here, the subscript \mbox{\footnotesize{B}} means that we have a
pairing wavefunction for composite {\bf B}osons.  This form suggests
more that $g_\tB(z_i,w_j)$ represents pairing of $p_x + i p_y$ type
since a phase of $+2 \pi$ is obtained when $z_i$ moves around $w_j$
rather than $-2 \pi$.   As suggested by
Ref.~\onlinecite{KimNayak01}, it seems natural to have the same
pairing symmetry for $d>\ell_B$ and $d\lesssim \ell_B$. This then
suggests that the relevant pairing symmetry for the composite
fermions is $p_x + i p_y$ rather than $p_x - i p_y$.  We emphasize
that it is mostly just a matter of nomenclature whether we label the
111 state as having $p_x + i p_y$ symmetry or $p_x - i p_y$
symmetry.   This ambiguity is a reflection of the fact that one can
attach Jastrow factors to electrons to construct new particles.
Depending on how the Jastrow factors are attached, the pairing can
appear either $p_x + i p_y$ or $p_x - i p_y$.   What is crucial,
however, is that the wavefunction always picks up a phase of $+2
\pi$ when $z_i$ moves around $w_j$
--- a behavior identical to that of the $p_x + i p_y$ paired CF
phase.  This similarity of the 111 phase and the $p_x + i p_y$
paired CF phase is crucial in the next section.

\subsection{Mixed CF-CB state}
\label{sec:MixedCFCBTheory}

In section \ref{sec:TheoryPairedCF}, we establish the general expression
for an interlayer-paired CF state in the bilayer
(\ref{eq:AllBilayerPaired}) which we believe should yield
appropriate ground state wavefunctions for large $d/\ell_0$.
Furthermore, in section \ref{sec:111-state} we determine a way to
write the 111 (CB) state, which is exact at vanishingly small
$d/\ell_0$, as a paired state. Both these types of wavefunctions can
be written as determinants of pairing functions $g_\tF$ and $g_\tB$,
respectively. Now, following the ideas of Ref.~\onlinecite{PRLSimon03}, we
consider transitional wavefunctions that include both the physics
of the CFs and the physics of the CBs.    We propose the following
extremely simple generalized form
\begin{subequations}
\label{eq:mixedPsi}
\begin{equation}
\label{eq:mixedPsiDet} \Psi^\text{CF-CB}=\det[G(z_i,w_j)]
\end{equation}
with \beq \label{eq:gplus}  G(z_i,w_j) =  g_\tF(z_i, w_j; \{ g_k
\}) + c_\tB g_\tB(z_i, w_j), \eeq
\end{subequations}
where $c_\tB$ is an additional
variational parameter representing the relative number of CBs
versus CFs. Note that as above, $g_\tF$  is a function of the
variational parameters $\{ g_k \}$ which describe the shape of the
pairing wavefunction.

In section \ref{sec:numericalResults} and Appendix
\ref{sec:CFCBOnSphere} we will translate these wavefunctions onto
the spherical geometry for which we have performed detailed numerics.

To elucidate the meaning of this linear interpolation between
composite fermion and composite boson pairing functions, it is
useful to consider more carefully the physics of the fermion pairing
described by Eq.\ \ref{eq:AllBilayerPaired}. Each entry in the
matrix $g_\tF(z_i,w_j)$ is a sum of many terms (See Eq.\
\ref{eq:gsum}) with each term representing the filling of particles
$z_i$ and $w_j$ into a particular pair of CF orbitals (one in each
layer). Upon multiplying out the entire determinant, each term will
include precisely $N$ occupied CF orbitals, and as required by Pauli
exclusion, no orbital may be occupied more than once. Terms with
double occupation of the same orbital cancel out by antisymmetry of
the determinant, even for non-orthogonal basis functions $\phi_i$.
The amplitude that a particular orbital is occupied is determined by
the coefficients $g_\bk$ (Compare Eq.\ \ref{eq:BCS2}). Now, let us consider
instead the pairing function $G(z_i,w_j)$ which has both the
fermionic $g_\tF$ terms as well as the bosonic $g_\tB$ terms (See
Eq.~\ref{eq:gplus}). When we calculate the determinant in
Eq.~\ref{eq:mixedPsiDet}, each $G(z_i,w_j)$ will be the sum of a term
where the CB orbitals are filled for particles $z_i$ and $w_j$ (the
$g_\tB$ terms) and several terms where $z_i$ and $w_j$ instead fill
a pair of CF orbitals. When we multiply out the entire determinant
it results in a linear combination of all possible choices of
filling $M$ CF orbitals and $N-M$ CB orbitals. As with the case for
the paired CF wavefunction, the amplitude of different orbitals
being filled is determined by the coefficients $g_\bk$ for the
fermions and $c_\tB$ for the bosons.

With this reasoning, we can actually reconstruct the mixed CB-CF
wavefunctions from Ref.\ \onlinecite{PRLSimon03} as a special case of
Eq.\ \ref{eq:mixedPsi}.   To this end, let us fix $c_\tB$ to some
constant value, e.g. $c_\tB=1$, and for all other
variational parameters $g_\bk$ let us use a step function (analogous
to Eq.\ \ref{eq:CFL_asPairedState} where we represented the filled
Fermi sea as a paired state), but with a reduced Fermi-momentum
$(k_F)_\tF$: \beq \label{eq:CFCB_FilledShells} g_\bk=\left\{
\begin{array}{ll}
\infty,& |\bk| \leq (k_F)_\tF \\
0,& \text{otherwise}
\end{array}.
\right. \eeq Where a very large $g_\bk$ is chosen, the corresponding
state is forced to be occupied (the resulting normalization suppresses anything
that does not include the maximal possible number of $g_\bk$ terms).
Due to the Pauli exclusion principle, every CF state may be occupied
only once, and consequently the particles remaining once the CF-sea
is filled up to the reduced Fermi momentum $(k_F)_\tF$ can only
occupy composite boson orbitals. The choice
(\ref{eq:CFCB_FilledShells}) results in the probability for a
CF to occupy a state with $|k| \leq (k_F)_\tF$ to be equal to unity,
which corresponds to a filled shell configuration. This construction
is ``equal" to the mixed CF-CB construction from Ref.~\onlinecite{PRLSimon03}.
(By ``equal" here we mean that the two constructions are
equivalent up to the differences between projection prescriptions in
the original Jain construction Eq.\ \ref{eq:simpleCFs} and the
Jain-Kamilla construction Eq.\ \ref{eq:Cf2}).   In Appendix
\ref{sec:AppFilledShells}, we show explicitly that the filled shell
states among those analyzed in Ref.~\onlinecite{PRLSimon03} can be reproduced
accurately by choosing $g_\bk$ as in Eq.\ \ref{eq:CFCB_FilledShells}.

It is very useful to remind the reader that both CFs and CBs could
in principle experience effective magnetic (or Chern-Simons) fields
due to their attachment to Jastrow factors.   As in
Ref.~\onlinecite{PRLSimon03}, we can write expressions for the
effective magnetic field ${\cal B}^\sigma$ seen by fermions ($\tF$) or
bosons ($\tB$) in layer $\sigma = \uparrow$ or $\downarrow$ as
\begin{eqnarray} \label{eq:Beff}
  {\cal B}^{\sigma}_\tF &=& B - 2 \phi_0 \, \rho_\tF^\sigma  - \phi_0 \rho_\tB \\
  {\cal B}^{\sigma}_\tB &=& B -  \phi_0 \, \rho \label{eq:Beff2}
\end{eqnarray}
where $B$ is is the external magnetic field, $\phi_0$ is the flux
quantum, $\rho = \rho^\uparrow + \rho^\downarrow$ is the total
density in both layers combined, $\rho_\tF^\sigma$ is the density of
CFs in layer $\sigma$ and $\rho_\tB = \rho_\tB^\uparrow +
\rho_\tB^\downarrow$ is the density of CBs in both layers combined.
It is important to note that precisely at $\nu=1/2 + 1/2$,
independent of the relative densities of CBs and CFs (so long as it
is symmetric between layers), at mean-field level, both species
experience zero total magnetic field.   For the mixed CF-CB state
with CF pairing, the number of CFs present may be uncertain.  As
mentioned above in the introduction, a pair of CFs in opposite
layers can transform into a pair of CBs in opposite layers.  It is
easy to see from Eqs. \ref{eq:Beff} and \ref{eq:Beff2} that this
process leaves the effective field seen by all species unchanged.

In contrast to the formula for the mixed fluid states given in
Ref.~\onlinecite{PRLSimon03}, the present form (Eq.~\ref{eq:mixedPsi}) with
$g_\bk$ according to Eq.~\ref{eq:CFCB_FilledShells} allows for
an efficient numerical calculation. In our present approach,
as explained below, the antisymmetry of the wavefunction is
a natural result of the determinant (requiring $\propto N_1^3$
numerical operations), whereas the wavefunctions from
Ref.~\onlinecite{PRLSimon03}  require explicit antisymmetrization,
an operation that requires much computation power with an
operation count scaling exponentially 
with the system size.

We emphasize again that while Ref.~\onlinecite{PRLSimon03} considered a limited
family of wavefunctions without CF pairing, the current approach
(Eq.\ \ref{eq:mixedPsi}) allows for the handling of both nontrivial CF
pairing and CF-CB mixtures simultaneously.

We now focus upon the question of whether, or under which
circumstances, Eq.\ \ref{eq:mixedPsi} is a valid lowest Landau level
wavefunction. First, to test the requirement of antisymmetry,
consider the interchange of 2 particles in the same layer, e.g.\ $z_i
\leftrightarrow z_j$, thus in all columns $k$:
\begin{subequations}
\begin{align}
\label{eq:symmetry_of_Psi}
\left\{
\begin{array}{ll}
g_\tB(z_i,w_k) \leftrightarrow g_\tB(z_j,w_k),& \text{ rows } i,j \\
g_\tB(z_l,w_k) \rightarrow g_\tB(z_l,w_k),& \forall\text{ rows } l \notin \{i,j\}
\end{array}
\right.\\
\left\{
\begin{array}{ll}
g_\tF(z_i,w_k) \leftrightarrow g_\tF(z_j,w_k),& \text{ rows } i,j \\
g_\tF(z_l,w_k) \rightarrow g_\tF(z_l,w_k), & \forall\text{ rows } l \notin \{i,j\}
\end{array}
\right.
\end{align}
\end{subequations}
In other words, exchanging two particles amounts to interchanging two rows
of the matrix $(G)_{ij}$.

The second condition to be checked is whether the proposed
wavefunction is properly homogeneous, implying that it is an angular
momentum eigenstate as required for the ground state of any
rotationally invariant system.  This condition is known to be true
for both limiting cases --- the 111 and the paired CF states. For
it to remain true for the mixed CF-CB state, it is sufficient to
require that $(g_\tF)_{ij}$ and $(g_\tB)_{ij}$ be of identical order
in all variables.  To check this it is sufficient to count the order
(or number of zeros) that occur for a given variable in $g_{ij}$.
For example, let us choose to look at the variable $z_1$.  For
$i \neq 1$ we have $g_\tB(z_i,w_j) = J_i^{zw} J_j^{wz}/(z_i - w_j)$.
The variable $z_1$ occurs only inside of $J_j^{wz}$ and occurs only
once. Therefore, it is first order in $z_1$.  Similarly for $i \neq
1$, in $g_\tF(z_i, w_j) = g(z_i,w_j) J_i^{zz} J_j^{ww}$ the variable
$z_1$ occurs only inside of $J_i^{zz}$ and occurs only one time, so
that it is also first order. Let us now look at the
term $i=1$.  In this instance, we have $ g_\tB(z_1,w_j) = J_1^{zw} J_j^{wz}/(z_1
- w_j)$ which has $z_1$ occuring $N$ times in $J_1^{zw}$, once in
$J_j^{zw}$ and once in the denominator, resulting in a total order
$N$.    For $g_\tF(z_1, w_j) = g(z_1,w_j) J_1^{zz} J_j^{ww}$ there are
$N-1$ powers of $z_1$ in $J_1^{zz}$ and additional $l$ powers in
$g(z_1,w_j)$ if we have $l$-wave pairing (See Eq.~\ref{eq:lWavePairing}),
giving a total number of powers of $z_1$
equal to $N-1+l$.   Thus, in order for this to match the degree of
$g_\tB(z_1,w_j)$, we must choose $l=+1$ or $p_x + i p_y$ pairing of
the Fermions.   It is clear that choosing any other pairing symmetry
would result in a wavefunction that is nonhomogeneous (therefore not
an angular momentum eigenstate) upon mixing fermions with
bosons.  While we cannot rule out some first order phase transition
between some other pairing symmetry for the CFs and a coherent CB
phase, it appears to us that $p_x + i p_y$ is the only symmetry
compatible with coexistence of CBs and CFs.

\section{Numerical Results}
\label{sec:numericalResults} In this section, we present
a numerical study of the
variational wavefunctions discussed previously.  In
particular we focus upon Eq.\ \ref{eq:mixedPsi}, which includes
Eq.~\ref{eq:BCS_RealSpaceCF} as an important special case. 
As our trial wavefunctions are given as variational 
states, we first need to optimize the variational parameters
$(g_k,c\tB)$ to obtain the optimal trial state for each
layer separation $d$. Given an explicit expression for a trial 
wavefunction at layer separation $d$,
Monte-Carlo may be used to numerically evaluate observables
such as the ground state energy, which we compare to similar
results calculated using exact diagonalization methods. We also 
evaluate the overlap of the trial states (\ref{eq:mixedPsi}) with
the exact groundstate wavefunctions.  We find
that our trial wavefunctions provide extremely accurate
representations of the exact ground states.

To avoid complications associated with system boundaries, except in 
section \ref{sec:pairingtopology} below, we choose always 
to work with the spherical geometry\cite{PRLHaldane83}
with a monopole of $N_\phi \equiv 2q$ flux quanta
at its center.  We give each electron not only a positional
coordinate, but also a layer index which may be either $\uparrow$ or
$\downarrow$.  $N$ electrons are put on the surface of the sphere
where half of them occupy each layer ($N = 2 N_1 = 2 N_\uparrow = 2
N_\downarrow$). We assume the limit of no tunnelling between the two
layers, therefore, these can be thought of as distinguishable electrons.  
We focus upon filling fraction $\nu=\half+\half$ which
corresponds to $N_\phi=2N_1-1 = N-1$.  This is precisely the
flux at which the 111 state occurs.  Note, however, that for a
single layer the composite Fermion liquid state with no effective
flux occurs at $N_\phi=2(N_1-1)$, which differs from what we
consider by a single flux quantum.  This difference in ``shift"
means that we are actually considering a crossover from the 111
state to a Fermi liquid state with one additional flux quantum.  It
turns out that this one additional flux quantum is appropriate here
since precisely such a shift is induced by the $l=+1$ nature of the 
$p$-wave pairing (that we determined as the appropriate pairing
channel in an earlier publication \cite{USPRL}).

On the sphere, the explicit form of the trial wavefunctions 
(\ref{eq:mixedPsi}) is defined by the expansion of the pair
wavefunction (\ref{eq:gsum}), where the basis functions $\phi_k$
become the monopole harmonics $Y_{q,n,m}$, with $q=\frac{1}{2}$ 
corresponding to $p$-wave pairing, as explained in detail in 
Appendix \ref{sec:CFCBOnSphere}. 

The interaction between electrons is taken to be the Coulomb potential
\begin{eqnarray}
    V_{\uparrow \uparrow}(r) &=& V_{\downarrow \downarrow}(r) =
e^2\, [\epsilon r]^{-1} \\
    V_{\uparrow \downarrow}(r) &=& V_{\downarrow \uparrow}(r) =
e^2\,\left[\epsilon \sqrt{r^2 + d^2}\right]^{-1}
\end{eqnarray}
where $r$ is the chord distance between the electrons, $\epsilon$
is a dielectric constant, and $d$ represents the distance between
the layers (measured in units of the magnetic length $\ell_0$).
Note that for simplicity, finite well width is not taken into
account.

Since our Hamiltonian is rotationally symmetric on the sphere, we
can decompose all states into angular momentum eigenstates. Our
exact diagonalization calculations determine the ground state to be
in the angular momentum $L=0$ sector. The trial ground state
wavefunctions are also $L=0$.   In addition to rotational symmetry,
the Hamiltonian exhibits a symmetry under exchange of the two
layers.   The ground state is found in the subspace with parity
$(-1)^{N_1}$. Again, it is simple to check that this is also the 
symmetry of our trial wavefunctions.

Exact diagonalization calculations are performed here for system sizes
of $N=10, 12\text{ and }14$ electrons for a large range of values
of the interlayer spacings $d$.   In order to evaluate the
significance of our results it is useful to examine the size of
the Hilbert space in which the Hamiltonian resides. While the full
Hilbert space is very large (even for 10 electrons), once the space
is reduced to states of $L=0$, the space is significantly smaller.
In Table \ref{tab:dimensions} we show the dimensions of the $L=0$
Hilbert space (and the dimensions of the even and odd parity parts
of that space) for the different size systems.  While these sizes
may appear small we note that they are typical sizes for
$L=0$ subspaces for what are considered to be relatively large
exact diagonalizations.   For comparison in Table
\ref{tab:dimensions} we show the dimensions of the $L=0$ spaces
for a number of other typical quantum Hall calculations in the
literature.

\begin{table}
\begin{center}
\begin{tabular}{c|c|c|c|c|c}
$\nu$ & $N$ & $N_\phi$ & $D(L=0)$ & $[E_\text{trial}-E_G]/E_G$ & $|\langle \Psi_\text{trial}|\Psi_G\rangle|^2$ \\
\hline
   & 5+5 & 9 & {\bf  29}+9 & $<1.4\times 10^{-3} $ & $ > 0.984(4) $ \\
$\half+\half$ & 6+6 & 11 & 97 + {\bf 155} & $<1.9\times 10^{-3}$  & $ > 0.978(4) $\\
  & 7+7 & 13 & {\bf 884}+715 & $< 2.2\times 10^{-3}$  & $ > 0.965(9) $\\
\hline
 & 6 & 15 & 6 & $5\times 10^{-4}$  & $ 0.99289 $\\
 & 7 & 18 & 10 &  $5\times 10^{-4}$ &$0.99273 $\\
$\frac{1}{3}$ &8&21&31 & $5\times 10^{-4}$ & $0.99082$ \\
 & 9 & 24 & 84 & $5\times 10^{-4}$ & $0.98816  $ \\
 & 10 & 27 & 319 & $6\times 10^{-4}$ & $0.984(3)$ \\
 & 11 & 30 & 1160 & $7\times 10^{-4}$ & $0.984(2)$ \\
\hline
 &8&16&8 & $4\times 10^{-5}$ & 0.9987(2) \\
$\frac{2}{5}$ & 10 & 21 & 52 & $2\times 10^{-4}$ & $0.9955(7)$ \\
 &12&26 & 418 & $2\times 10^{-4}$ & $0.994(2)$\\
\hline
\hline
\end{tabular}
\end{center} \caption{ \label{tab:dimensions} Hilbert space dimensions of the
$L=0$ subspace for the examined bilayer systems and several reference
states. For bilayer states two values are indicated corresponding
to the fraction of states with odd and even parity under layer
exchange. The respective subspace containing the ground state is
typeset in bold. Data on the exact energies of single layer states was
collected from [\onlinecite{RegnaultHome}]. The last column indicates
overlaps of the respectively appropriate trial wavefunctions with the
exact ground states (data from Ref.~\onlinecite{YoshiokaBook} or from our calculations,
where errors are indicated).}
\end{table}

For a given interlayer spacing $d$, we first perform exact
diagonalization to find the ground state, and then determine how
``close" we can get to this state with a variational
wavefunction.  The variational wavefunction is a function
of the parameters $\{ g_\bk \} $ (for both  Eq.\ \ref{eq:mixedPsi}
and Eq.\ \ref{eq:BCS_RealSpaceCF}) and one additional parameter
$c_\tB$ (which we can think of as being set to zero in
Eq.~\ref{eq:BCS_RealSpaceCF}).  While it is clear that with enough
variational parameters one can fit any result, the actual number of
variational parameters we use is quite small. First of all $g_\bk$
can be assumed to be a function of $|\bk|$ only.   More accurately,
on the sphere the orbital states are indexed by the quantum numbers
$n$  (the shell index) and $m$ (the $z$ component of the angular
momentum in the shell), and by rotational invariance of the ground
state we can assume that the variational parameters are independent
of $m$ (as detailed in Appendix \ref{sec:CFCBOnSphere}).  
In other words, there is a single parameter per composite
fermion shell (or composite fermion Landau level);  we label these
parameters as $g_n$. For the system sizes available in our exact
diagonalizations, no more than 5 such variational parameters are
necessary to obtain satisfactory trial states. Considering the
dimensions of the symmetry reduced Hilbert space (shown in Table
\ref{tab:dimensions}) which is much larger than 5, we conclude that
the agreement of our states with the exact ground state is
nontrivial.

There are several ways to evaluate the quality of a given trial 
wavefunction (or the ``closeness" of a trial wavefunction to an 
exact wavefunction).   For example, one could compare the energy 
of the trial wavefunction to that of the exact ground state energy. 
By the variational principle, if one obtains the exact ground state 
energy, then the trial wavefunction must be the exact ground state. 
Another well-known measure of the quality of a trial wavefunction is 
the overlap of the trial wavefunctions with the exact ground state. 
We shall adopt these two measures of accuracy for the analysis in 
the main text of the current paper.

The details of the optimization methods used to obtain the right
variational parameters for a good trial state at a given layer 
separation $d$ are explained in Appendix \ref{sec:numericalmethods}.
In brief, however, we proceed as follows. If we optimize for the
ground state energy $E$, a Monte-Carlo estimate of the Hamiltonian 
operator $\langle H(d)\rangle$ is obtained in a very restricted basis of states defined
by the trial wavefunction $\Psi_0$ to be studied, and an initial guess of
variational parameters $\vec\gamma=\{c_B,g_0, g_1,\ldots\}$. This basis is spanned
by $\Psi_0$ and its derivatives $\Psi_n=\partial\Psi_n/\partial g_n$ with 
respect to $g_n$. Diagonalizing the estimator $\langle H(d)\rangle$ yields
a new set of variational parameters, which are used as an improved guess of
$\vec\gamma$. This procedure is iterated until convergence is reached. If we optimize
for the overlap with the exact ground state wavefunction, the procedure
is simpler as we can directly evaluate the gradient of the overlap 
$\partial/(\partial \gamma_i) |\langle\Psi_\text{trial}|\Psi_\text{exact}\rangle|^2$.
Updating $\vec\gamma$ according to a steepest descent algorithm has proven 
sufficient to optimize the overlap. For further details, please refer to 
Appendix \ref{sec:numericalmethods}.

In addition to the energy and the overlap with the exact ground state, one could compare the pair correlation functions (both inter-layer and intra-layer) of the trial wavefunction to that of the exact ground state. Since, for pairwise interactions, the pair correlation function completely determines the energy of the system, again, a trial wavefunction that has the exact ground pair correlation function must identically be the correct ground state. Such a comparison of correlation functions is given in Appendix \ref{sec:correlations}.

For very large system sizes of course we are unable to perform exact
diagonalization.  Nonetheless,  we are still able to study this system
by Monte-Carlo.  In such cases, the variational parameters
are optimized by simply attempting to minimize the energy of the trial
state (as discussed in Appendix \ref{sec:numericalmethods}), 
though we are uncertain of the proximity of the results to
the exact ground state. At present, this possibility has not yet been
fully exploited, and we limit our study of bigger systems to filled
shell states. This study is presented in Appendix
\ref{sec:AppFilledShells}.

\subsection{Paired CF results}
\label{sec:pairedResults} 
\begin{figure}[ttbp]
  \begin{center}
    \psfrag{ylabel}[c][t][1.5]{$|\langle \Psi^\text{CF-BCS}|\Psi_\text{exact}\rangle|^2$}
    \psfrag{xlabel}[c][b][1.5]{$d\;[\ell_0]$}
    \includegraphics[width=0.9\columnwidth]{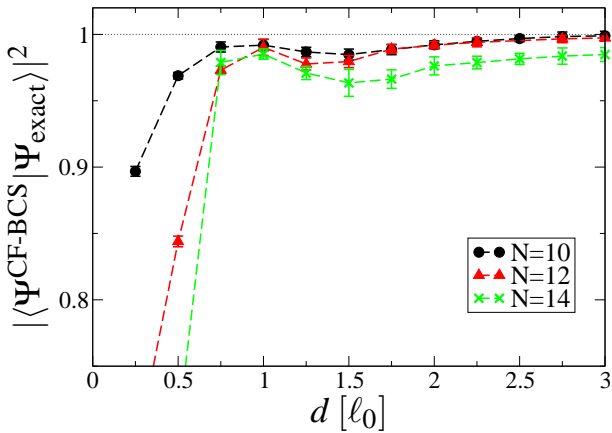}
  \end{center}
\caption{  \label{fig:overlapsCFonly}  Squared overlaps of $(p_x + i p_y)$-wave paired CF trial states
(Eq.~\ref{eq:AllBilayerPaired}) with the exact ground state, for $N=5+5$, $6+6$ and $7+7$. For $d \gtrsim \ell_0$ extremely high overlaps are obtained.  However, for $d \lesssim \ell_0$ the CF-BCS trial wavefunctions are not accurate, suggesting a phase transition around $d \approx \ell_0$.}
\end{figure}

\begin{figure}[ttbp]
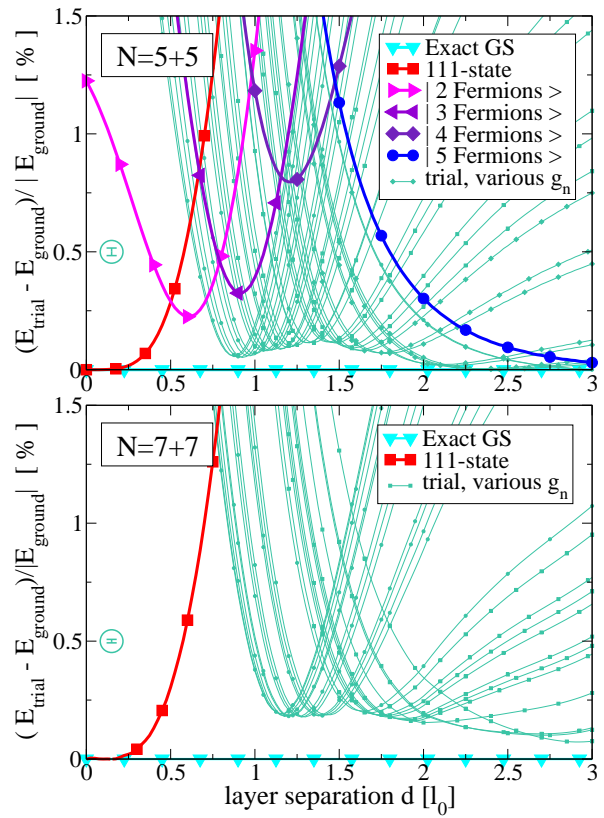
 
  \begin{center}
    \includegraphics[width=0.9\columnwidth]{GS_CFpairing_noCB_NCF.eps}
    \includegraphics[width=0.9\columnwidth]{energies_N7_noCB_pc.eps}
  \end{center}
\caption{  \label{fig:EnergiesCFP}  Relative errors in
energy of $(p_x + i p_y)$-wave paired CF trial states
(Eq.~\ref{eq:AllBilayerPaired}) in the bilayer for $N=5+5$ particles
(top) and $N=7+7$ particles (bottom). Each of the represented curves
corresponds to a different trial state, i.e.\ a different choice of
parameters $\{g_n\}$.  The vertical axis is the fractional energy
difference of the trail wavefunction energy with resepect to the
exact ground state energy  $(E_\text{trial} - E_\text{ground})/E_\text{ground}$.
The largest errors are of order $1.4\times 10^{-3}$ and $2.2\times
10^{-3}$ for $N=5+5$ and $N=7+7$ respectively, when regarding only
those layer separations greater than $d=d^\text{\tiny CB}$, where
the paired CF Ansatz yields ``good'' trial states. The encircled error
bar indicates the magnitude of Monte-Carlo error. For comparison,
the mixed fluid trial states from Ref.~\onlinecite{PRLSimon03} are
represented as bold lines in the upper panel (see legend). }
\end{figure}

In this section, we discuss the results
for the paired CF wavefunctions (\ref{eq:AllBilayerPaired}) with
pairing in the $p_x+ip_y$ channel.   Figure \ref{fig:overlapsCFonly} 
shows overlaps of our trial states with the exact ground state for 
several system sizes as a function of interlayer spacing.  (This data 
has been previously presented in Ref.~\onlinecite{USPRL}). In Figure \ref{fig:EnergiesCFP},
the relative errors of the trial state energies $E_\text{trial}$ with
respect to the ground state energy $E_G$ are represented
as $[E_\text{trial}(d,\{g_n\})-E_G(d)]/E_G(d)$
for two different system sizes of $N=10$ and $N=14$ particles.
From these two figures, it is clear that the paired CF states yield excellent trial states for large $d$, whereas there is a layer separation $d^\text{\tiny CB}$
below which the paired CF picture yields no good trial states. We
find $d^\text{\tiny CB}\approx 0.9\ell_0$ and $d^\text{\tiny
CB}\approx 1.1\ell_0$ for 10 and 14 particles respectively.
For 12 electrons (not displayed), this value amounts to
$d^\text{\tiny CB}\approx \ell_0$ (See also Table \ref{tab:scaling}).

These results are surprising, since the regime where paired CF
states yield very good trial states extends from infinite layer
separation down to $d\sim \ell_0$, well below the point where
experiments observe the set-in of the various phenomena that are
thought to be associated with spontaneous interlayer coherence and
the presence of CBs or interlayer excitons. Given the large increase
in $d^\text{\tiny CB}$ between the systems with $N=10$ and $N=14$
particles, it is not clear at present how this extends to larger
systems. A na\"ive linear extrapolation with respect to the inverse
system size $N^{-1}$ based on the above values yields $d^\text{\tiny
CB} \approx 1.76$ in the thermodynamic limit, which is rather close
to where a transition is observed experimentally.

Unfortunately extrapolation to the thermodynamic limit is made difficult 
by shell filling effects.   In particular, $N=12$ corresponds to having two CF
shells filled in each layer (the lowest shell has two electrons per
layer, and the next shell has 4 electrons per layer. See Appendix
\ref{sec:AppFilledShells}).   Thus, this particular system size could behave differently from the $N=10$ ($N=14$) case, where there is one CF-hole(electron) in the valence shell in each layer.   Indeed, at large $d$, shell filling effects are quite strong, as was discussed in depth in Ref.~\onlinecite{USPRL}.  In particular, it was found that for large enough $d$, the system always follows Hund's rule,\cite{ReadRezayiCFFL} maximizing the angular momentum of the valence shell within each layer.   Only for system sizes with filled shells (such as $N=12$), or when there is a single electron or single hole in the valence shell in each layer (such as $N=10$ and $N=14$) can the Hund's rule state be expressed as a CF-BCS wavefunction  in the form of (\ref{eq:AllBilayerPaired}). For other cases, the large $d$ limit of the CF-BCS states differs from the Hund's rule state. However, as argued in Ref.~\onlinecite{USPRL}, this Hund's rule physics, involving only the $\sqrt{N}$ particles in the valence CF shell, should become less important as one goes to larger and larger systems.   If one assumes that the energy gain of pairing is roughly $\Delta N$ as is usual for BCS theory, then for any finite $\Delta$, the pairing energy gain will always be larger than any putative Hund's rule energy gain in the thermodynamic limit.

In Ref.~\onlinecite{USPRL} arguments and detailed numerics were given supporting this picture: that for  large $d$ in the thermodynamic limit, the CF-BCS state prevails over the Hund's rule state.  However, for very large $d$, with very weak coupling between the layers, no definite numerical conclusion could be reached.  Nonetheless, whether or not one can draw conclusions about very large $d$, it is certainly the case that the numerics {\it strongly} suggested the existence of a CF-BCS phase for a range of intermediate $d$ where the Hund's rule physics is not present.

For simplicity, in this paper, since we are concerned mostly with the physics at smaller $d$ (and where an incompressible quantum liquid is observed), we will not address the Hund's rule physics further.   To avoid this complication, we will focus on shell fillings such that Hund's rule is compatible with the CF-BCS state, so no competition arises.   We refer the reader to Ref.~\onlinecite{USPRL} for further discussion of this issue.

\subsection{Mixed CF-CB results}

\label{sec:mixedResults} In order to obtain a complete description
of the ground-state  for small layer spacing $d$, we need to
consider the mixed fluid description of the quantum Hall bilayer.
Upon addition of CBs to the paired CF description, one obtains the
family of mixed CF-CB states (Eq.~\ref{eq:mixedPsi}). Technically
this corresponds to adding one more variational parameter to the
previously discussed case of paired CFs. Consequently, using this
extended family of trial states yields at least as good results as
with composite fermions only.

\begin{figure}[ttbp]
  \begin{center}
    \psfrag{ylabel}[c][t][1.5]{$|\langle \Psi^\text{CF-CB}|\Psi_\text{exact}\rangle|^2$}
    \psfrag{xlabel}[c][b][1.5]{$d\;[\ell_0]$}
    \includegraphics[width=0.9\columnwidth]{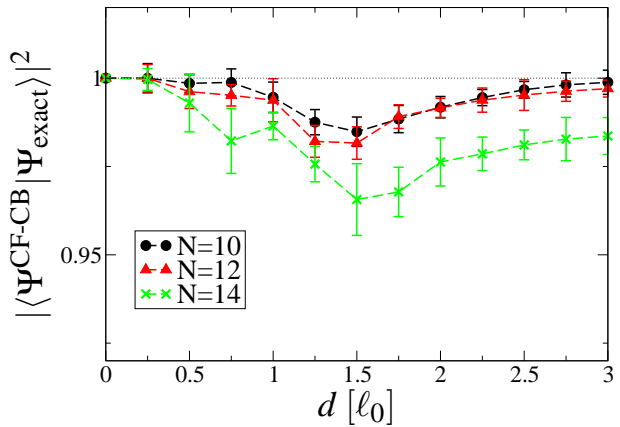}
  \end{center}
\caption{  \label{fig:overlapsCFCB}  Squared overlaps of the exact ground state with trial state for mixed CB-CF fluid with interlayer $(p_x + i p_y)$-wave
pairing.  Over the entire range of $d$, extremely high overlaps are obtained.  Data is shown for $N=5+5$, $N=6+6$ and $N=7+7$.  The quality of the overlaps are comparable to that of the Laughlin state, see Table \ref{tab:dimensions}.}
\end{figure}

\begin{figure}[ttbp]
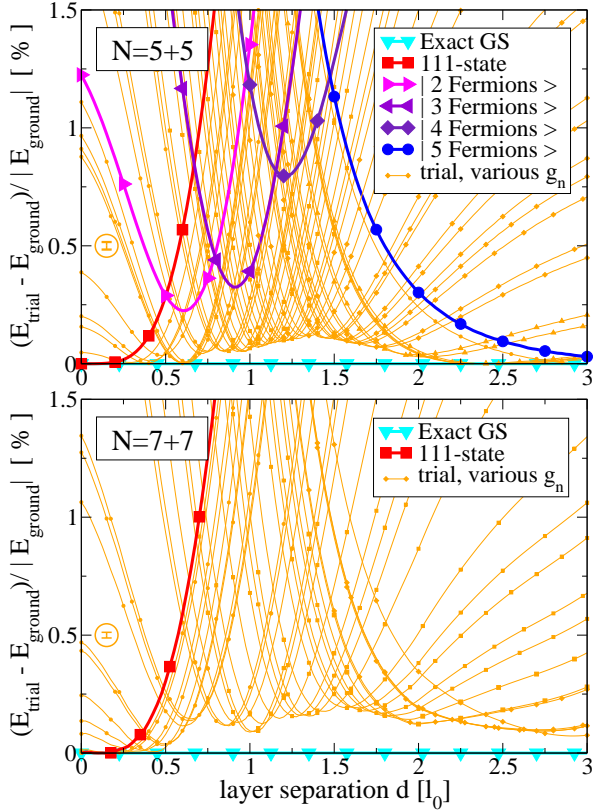

  \begin{center}
    \includegraphics[width=0.9\columnwidth]{GS_CF-CB_NCF}
    \includegraphics[width=0.9\columnwidth]{energies_N7_CF-CB_pc.eps}
  \end{center}
\caption{  \label{fig:EnergiesMixed}  Comparison of relative errors
in energy for mixed CB-CF fluid with interlayer $(p_x + i p_y)$-wave
paired CF model systems with $N=10$ (top) and $N=14$ electrons
(bottom). As in Fig.\ \ref{fig:EnergiesCFP}, each curve represents a
different trial state. The mixed fluid states from
Ref.~\onlinecite{PRLSimon03} are highlighted in bold in the upper
panel. Over the entire range of $d$, extremely good trial states are
obtained, with a remaining error $\delta\epsilon < 2.2\times
10^{-3}$. For intermediate $d$, where the states without pairing $\state{n}$ do not
perform very well, considerable improvements are realized. Monte-Carlo
errors are on the order of the encircled error bar.}
\end{figure}

Numerical simulations confirm that the mixed fluid description of
bilayer trial wavefunctions (Eq.~\ref{eq:mixedPsi}) achieves an
impressively precise description of the ground state for all $d$.
This is borne out by the numerical results shown in Fig.~\ref{fig:overlapsCFCB} and Fig.~\ref{fig:EnergiesMixed}, analogous to the above Figs.~\ref{fig:overlapsCFonly} and \ref{fig:EnergiesCFP} except that now we have used the mixed fluid wavefunctions.

In Fig.~\ref{fig:overlapsCFCB} we find that over the entire range of $d$, the overlap with the exact ground state is extremely high for all systems sizes.   The lowest overlaps occur at roughly $d=1.5 \ell_0$.   As seen in Table \ref{tab:dimensions} these ``worst case" overlaps are comparable to the overlaps seen for the Laughlin $\nu=1/3$ state for Hilbert spaces of similar size.  Writing squared overlaps as $1 - \delta$, we find that the $\delta$ value for our worst trial wavefunctions are roughly twice that of the Laughlin state for similar Hilbert-spaces of comparable dimension.  Similarly, in Fig.~\ref{fig:EnergiesMixed}, we find that the largest relative error for the prediction of the ground-state energy occurs at intermediate distances
close to $d=1.5\ell_0$.   These ``worst case errors'' are also listed in
Table \ref{tab:dimensions}.  We find that the energy errors for our bilayer states are about 3-4 times as large as those of the Laughlin state at $\nu=1/3$ for
Hilbert-spaces of comparable dimension.   Given that the Laughlin
state is often referenced as a ``gold-standard" for its accurate description
of the exact ground state, we find the level of accuracy of our trial states to be quite satisfactory.   (Note that the CF
wavefunctions for $\nu=2/5$ are even more accurate than the Laughlin
state at $\nu=1/3$ for comparable Hilbert-space dimension).  At layer separations $d$ not too close to $1.5\ell_0$, the
bilayer trial wavefunctions are even more accurate than the number
quoted above, and may exceed the
accuracy of the Laughlin and even of the $\nu=2/5$ trial wavefunction.

For further comparison, in the upper frame of Fig.~\ref{fig:EnergiesMixed} are the energies (dark lines) of the mixed fluid wavefunctions first introduced in Ref.~\onlinecite{PRLSimon03}. As discussed above, these wavefunctions lack CF pairing that is included in Eqs.~(\ref{eq:AllBilayerPaired}) and (\ref{eq:mixedPsi}).   Although these wavefunctions clearly capture some of the physics of the crossover from the 111 to the CF liquid, it is clear that pairing is required in order to have a high degree of accuracy. 

Naturally, nearly exact trial states are obtained at $d\to 0$, where
the appearance of CFs may be regarded as a perturbation of the
111-state (which is obtained by the particular choice of parameters
$g_k=0$ and $c_\tB=1$, and which is the exact ground state at
$d=0$). However, the admixture of CFs becomes important at rather
small $d$.   We see that this admixture provides a nearly exact description of the fluctuations around the 111 state.  However, in the regime of small layer separation, an equivalent description in terms of other excitations to the 111-state may be
also suitable.\cite{JoglekarPRB01,Milica}

It should be noted that the number of variational parameters
required to obtain good trial states becomes maximal at
intermediate layer separations $d\sim 1.5 \ell_0$.  However,
even at $d \sim 1.5 \ell_0$, only four variational parameters 
are required for the system sizes we consider. In the limits
of $d=0$ and $d\to\infty$, writing the wavefunction in the form
(Eq.~\ref{eq:mixedPsi}) essentially amounts to rephrasing a
parameterless trial state, respectively the 111-state and a product state of 
Fermi-liquids, in a different form (the case of $d\to\infty$
is slightly more complicated, as was discussed in detail in Ref.~\onlinecite{USPRL}).
As either regime is approached, the number of variational parameters 
required to describe the physics of the ground state decreases.
For example, at $d \sim 0.5 \ell_0$ and $d \sim 3 \ell_0$ only two variational 
parameters are used.

It is at intermediate distances of $\ell_0\lesssim d \lesssim 2\ell_0$
that the mixed fluid state is most different from both the 111-state 
and the CF liquid. In this regime, the influence of CF pairing is strongest, 
and the CFs tend to occupy orbitals in CF shells higher than the Fermi momentum 
of a filled CF Fermi sea (as shown in section \ref{sec:occupation}, below).
Although the overlap of our trial states has a minimum seen at $d\approx 1.5\ell_0$,
which occurs in the regime that we identify as a paired state, we would like to 
point out that the magnitude of this overlap remains very high. In fact, 
the overlap is larger than that found for paired states in
the single layer, i.e., for the Moore-Read state, \cite{MooreRead91} or
its generalizations for trial states in the weak-pairing phase at 
$\nu=5/2$,\cite{USSingleLayer} which is accepted to describe the
physics of the quantum Hall state at that filling factor.
Similarly, we conclude that CF pairing captures the essential physics
of the quantum Hall bilayer system at filling factor $\nu=1$ for intermediate
layer separations $d$.

As a side note, we have confirmed numerically that the mixed fluid trial states
from Ref.~\onlinecite{PRLSimon03} may be obtained in a manner prescribed
in the approach to the filled shell cases. The general
phenomenology that may be obtained from the analysis of filled CF
shell states is discussed in Appendix \ref{sec:AppFilledShells}.

\subsection{Occupation probabilities of CF shells}
\label{sec:occupation}
With the mixed fluid wavefunctions (Eq.~\ref{eq:mixedPsi}), a vast
family of trial states is available. Furthermore, the above results
confirm that the mixed fluid wavefunctions allow for an accurate
description of ground-state properties. As a step towards an
understanding of the numerical results just presented, it is
interesting to characterize the most successful trial states
via the probability for an electron to occupy a given CF-LL
within such a state.

In Figures \ref{fig:overlapsCFonly}-\ref{fig:EnergiesMixed}, the
various trial states were shown without specifying the explicit
values of the variational parameters $\{g_n\}$.\cite{NoteAdditionalMaterial}
Indeed, giving the precise values of these parameters may likely not have 
been very meaningful to the reader for two reasons. First, these parameters
are defined only up to an overall global normalization. Secondly,
and more importantly,
the normalization of the individual composite fermion orbitals that
the wavefunction is composed of is not well defined. If a basis of 
normalized single particle orbital is projected to the LLL using 
Eq.~\ref{eq:tildephi}, we obtain a basis of many-body composite fermion 
orbitals that are no longer orthogonal, and which have lost their
original normalization. 
In particular, the projected orbitals $\tilde\phi_i=\tilde\phi_i(z_1,\ldots,z_N)$ 
become functions of all particles' coordinates. Their normalization 
$\mathcal{\tilde N}$ could be defined by integrating out all coordinates but one.
However, in such a definition, the normalization $\mathcal{\tilde N}$
of a single orbital becomes ill defined, as it also depends on the correlations
in the system, which however, are only known after a complete many-body state
has been specified.

Since the normalization of the orbitals we use is ill defined,\cite{RefAdditionalMaterial} 
we propose a universally applicable definition of 
the occupation $p(k)$ of a CF orbital $\tilde \phi_k$ with momentum $k$ to be given by
 \beq \label{eq:occupationProba} p(k) =
\frac{1}{2N}  \frac{\partial
 \log \langle \Psi(\{g_k\}) | \Psi(\{g_k\}) \rangle}{\partial \log g_k}, \eeq where  $\Psi(\{g_k\})$ is
the bilayer wavefunction which is a function of the variational parameters $g_k$, and
$\langle\cdot\rangle$ denotes the unnormalized Monte-Carlo average.
The relation (\ref{eq:occupationProba}) was successfully deployed
for pairing in a single layer by two of the current authors,\cite{USSingleLayer}
and may be explained with the example of a simple one-particle two-state model
with wavefunction $\Psi=g_1\phi_1 + g_2 \phi_2$, which we allow to
be unnormalized. Expanding the square of this wavefunction, \beq
\langle \Psi | \Psi \rangle  =\sum_{i,j=1,2}  g_i^*  g_j \, \, \langle \phi_i | \phi_j \rangle \eeq we can
see that Eq.\ (\ref{eq:occupationProba}) yields the
proper occupation probabilities of both levels, provided that the
overlap integral $\langle\phi_1 | \phi_2 \rangle$ vanishes. This
is the case for the scalar product of wavefunctions in a
regular orthogonal basis. This argument
generalizes to the many-body case simply by applying the product-rule for
the derivative.

For the mixed bilayer states, however, we use the non-orthogonal
basis of the LLL-projected CF orbitals. Nonetheless, we could verify
that the occupation probabilities for states with filled CF shells
(where we know the occupation probabilities (See Appendix
\ref{sec:AppFilledShells})) are obtained from
(\ref{eq:occupationProba}) with very high accuracy, showing that the
respective overlap integrals are small, thus giving a physical
meaning to these occupation probabilities.

Surprisingly, applying Eq. \ref{eq:occupationProba} to the variational
parameter for composite bosons $c_\tB$, does not yield the proper
value for the occupation probability of the CB orbital.
Consequently, we exploit the fact that this probability is complementary
to the total occupation probability of the various CF orbitals. This
allows to calculate the occupation probability of the CBs $p_\tB$ as
\beq \label{eq:bosonOccupation}
p_\tB=1-\sum_n p(n),
\eeq
where $p(n)$ is the probability to find an electron in CF shell $n$.

\begin{figure}[ttbp]
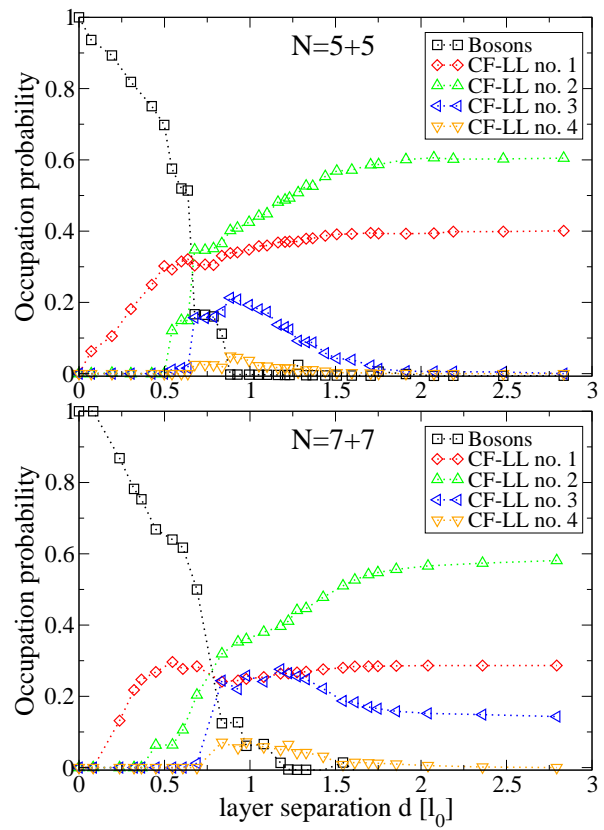

  \begin{center}
    \includegraphics[width=0.9\columnwidth]{OccupationsN5.eps}
    \includegraphics[width=0.9\columnwidth]{OccupationsN7.eps}
  \end{center}
\caption{  \label{fig:occupations}  Probabilities for a single
electron to occupy a given orbital, as obtained from
Eq.\ \ref{eq:occupationProba}. A region of strong pairing, i.e.\ large
probabilities to find an electron in an excited orbital above the
would be Fermi-momentum, is found between $d\sim 0.8 \ldots
1.5\ell_0$. Note that the probability $p_\tB$ that an electron forms a CB
(obtained as $p_\tB=1-\sum p(n)$) practically drops to zero,
or slightly below, at $d\sim \ell_0$. The kinks in the dependence of
$p_\tB(d)$ are close to values which are related to the CF shell
structure.}
\end{figure}

Let us now turn to the results obtained for the two selected systems
sizes that we discussed in the previous sections. Taking
the best trial state as a reference at each $d$, we may extract
from our calculation the approximate separation-dependence of
the occupation probability $p(n)$. The resulting data is displayed in
Fig.\ \ref{fig:occupations}.

We discuss these results going from right to left on the axis of layer
separations. Upon looking at large
layer separations, it is first noticed that the distribution at $d=3
\ell_0$ is that of the CF Fermi sea. For example, in the lower
panel for $N=7+7$ electrons, the probability that an electron is in
the lowest CF shell is $p(0)\approx 2/7 \approx 0.28$. For the next
higher shell, which is fourfold degenerate, one finds $p(1)\approx
4/7 \approx 0.57$. The third shell accounts for the remaining
probability. Upon going to intermediate layer separation, one notices
the onset of pairing as one would expect by analogy with BCS theory:
electrons are lifted above those orbitals within the equivalent of a
Fermi-sea and occupy states at higher momentum, instead.
Correspondingly, the occupations in the lowest two shells drop to
allow the occupation of the higher ones ($n=3$ included, which is occupied
by a single electron per layer, initially). For $N=5+5$, we follow an analogous
trend of redistribution among the occupation of CF-levels,
noting that the total probability of finding a particle in one of the
excited orbitals is quite large, with absolute values close to
25\%. Only at lower layer separation does the occupation of the CB
orbital become important. Conversely, the occupation of CF orbitals
plays an important role down to very low layer separations.

Now, the occupation of the CB orbital $p_\tB$ shall be analyzed. At
large layer separation, the value obtained from (\ref{eq:bosonOccupation})
drops slightly below zero. This is an inconsistency related to the
empirical character of Eq. \ref{eq:occupationProba}.
However, the error is not very large, amounting to about
$1\%$, which gives some confidence into our method, though it
reminds us that it is approximate. We need to remark also, that the
data is based on calculations for a restricted number of trial
states, such that more substantial deviations are likely due to
data that corresponds to not quite optimal
trial states. The roughness  of the curves
illustrates this. Some of the features in the behavior of
$p_\tB(d)$ might also be caused by filled shell (i.e., finite size) effects, given that kinks
are featured at values close to $1-n_S/N$ where $n_s$ CFs
yield a filled shell configuration.

\subsection{Order Parameters}
\label{sec:symmetries} This section is devoted to discussing another
means of characterizing the mixed fluid trial states --- we discuss
the broken symmetries of our wavefunctions and their associated
order parameters.   In the present case of the bilayer system with
paired CF, two distinct symmetries, will be discussed in sections
\ref{sec:exsupord} and \ref{sec:cfpairord}. In addition, we consider
an additional topological order parameter of the paired CF system in
section \ref{sec:pairingtopology}.

\subsubsection{Excitonic Superfluid Order}
\label{sec:exsupord}  In order to consider the first of the two
potential symmetries of our quantum states, it is useful to employ
the pseudospin picture. A density-balanced bilayer system has been
described as a pseudospin field with its values confined to the
$x$-$y$-plane.\cite{MoonetalPRB95} In the ground state the
orientation of this pseudospin field is homogeneous and (in the
absence of interlayer tunneling) a spontaneous breaking of
the U(1) symmetry for rotations of the pseudospin around the
$z$-axis occurs such as to select a preferred direction in the
$x$-$y$-plane. The operator for the in-plane pseudospin thus yields
a measure for detecting the symmetry of a coherent state in the
bilayer system. In second quantized notation, this order parameter
describes a flip of the pseudospin at position ${\bb r}$, noted as
$\mathcal{F}({\bb r})$:
\begin{equation}
  \label{eq:flip_operator}
  \mathcal{F}({\bb r})\equiv \Psi^\dagger_\uparrow(\mathbf{r}) \Psi_\downarrow({\bb r})
\end{equation}
For the purpose of numerics at fixed particle number $N_i$ per layer, the operator needs to be
modified such as to conserve $N_i$. This is realized by taking the product
$\mathcal{F}({\bb r})\mathcal{F}^\dagger({\bb r}')$ at two distant points ${\bb r}$ and ${\bb r}'$ which now preserves
the number of particles in each layer.
In the limit $|\bb r -\bb r'|\to\infty$, one expects to recover the square of the
expectation value of $\mathcal{F}$ in a corresponding grand canonical ensemble.
Thus we define
\begin{equation}
  \label{eq:shift_operator}
  {\mathcal S} = \lim_{|\bb r -\bb r'|\to\infty}\mathcal{F}(\bb r) \mathcal{F}^\dagger (\bb r')
\end{equation}
For a finite sized system, we must be content to move the positions $\vec r$ and $\vec r'$ as
far apart as possible.  One can visualize the action of this operator either as the
associated pseudospin flips of two electrons in opposite layers at
distant positions, or as the exchange of the real-space positions of
these two particles. This operator can be easily calculated in our
Monte-Carlo simulations carrying out this kind of exchange in position
for pairs of electrons and monitoring the effect on the wavefunction.

For the 111-state, we have $\bra{111}\mathcal{S}\ket{111}=-1$.
Conversely, $\bra{\text{CFL}}\mathcal{S}\ket{\text{CFL}}$ yields a
very small value provided that the distance $|\bb r -\bb r'|$ is
chosen to be sufficiently large. Any finite geometry imposes a constraint on the
limit in (\ref{eq:shift_operator}), but numerics confirm that
$\bra{\text{CFL}}\mathcal{S}\ket{\text{CFL}}\approx 0$ to within roughly a part in  $10^{-5}$
for accessible system sizes.
As the sign of $\langle  {\mathcal S} \rangle$ does not matter to distinguish the 111 and paired
CF phases, we will refer to its absolute value
\beq
S = |\langle  {\mathcal S} \rangle|
\eeq
as the excitonic superfluid order parameter.
Upon calculating $S$ for mixed fluid states
with filled CF-shells, we find that there is a monotonic relation
between the order parameter $S$ and the fraction
of electrons $N_\tB/N$ that have undergone a CB-like flux attachment
(See Appendix \ref{sec:AppFilledShells}).
Furthermore, results for several different system sizes collapse on
a single curve, such that we may estimate finite size effects to be
small. We conclude that $S$ is indeed a suitable order parameter
for the transition between the CFL and the 111-state.

\begin{figure}[ttb]
  \begin{center}
    \includegraphics[width=0.9\columnwidth]{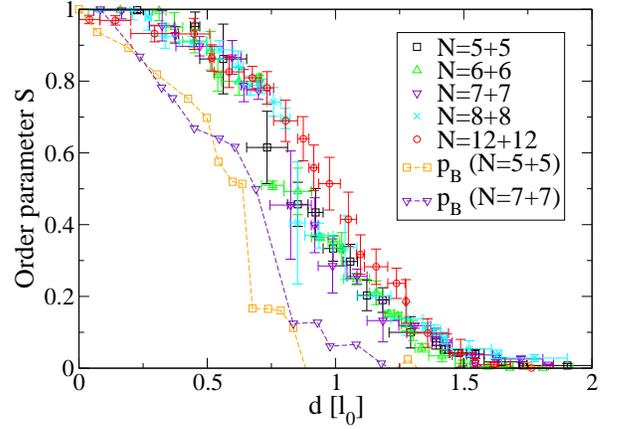}
  \end{center}
\caption{  \label{fig:orderParameter}  A plot of the excitonic superfluid order parameter
$S(d)$ for different system sizes according to the legend
(symbols with error bars) and the fraction of bosons $p_\tB$ as
obtained from Eq.\ \ref{eq:occupationProba} (dashed lines). Data of
system sizes $N=8+8$ and $N=12+12$ is based on a set of MC calculations 
with energy optimization (see Appendix \ref{sec:numericalmethods})
and we have no according exact calculations available
for comparison.}
\end{figure}

While it is true that increasing the fraction of CBs yields a larger
order parameter, this is not the only factor influencing $S$.
In particular, for our finite sized systems, nonzero values of the order parameter can be
obtained for bilayer states within the paired CF picture, i.e.\
\emph{without} composite bosons.  Let us discuss this feature in detail by examining
$S$ calculated in our Monte-Carlo
simulations for each of our trial states.   We attribute the value obtained for the best trial
state at a given $d$ to represent the value $S$ in the ground state at that $d$ to a very good approximation.
The data in Fig.~\ref{fig:orderParameter} was obtained
following this procedure. Error bars are established by taking into
account the values of $S$ for trial states, whose energies
are within the range of Monte-Carlo errors from the best trial state.

A non-zero excitonic superfluid order parameter (i.e., $S$) for
pure paired-CF states means that
good trial states without adding composite bosons can be found above
some layer separation $d^\text{\tiny CB}$ which is well below the
value $d_c$, where $S$ becomes non-zero.
While it is not easy to determine exactly the layer separation where
mixed CB-CF fluid states become substantially better than the pure paired-CF states,
it is more straightforward to estimate the paired CF states' maximal possible
order parameter
\beq
S_\text{max}=\max_{\{g_k\}} \left|\langle \Psi^\text{CF-BCS}_{\{g_k\}} | \mathcal{S}
| \Psi^\text{CF-BCS}_{\{g_k\}}\rangle\right|,
\eeq
where the maximization is over only paired CF states without any CBs.  The fact that $S$
can be nonzero without CBs is itself an intriguing phenomenon.
For instance, considering $N=5+5$ electrons, $p$-wave paired CF states
yield a maximum $S_\text{max}$ as large as 42\% the
value of the CB condensate (the 111 state). Values for other system sizes are given in
Table \ref{tab:scaling}. The numbers indicated for $S_\text{max}$
should be understood as estimates of a lower limit of this value. They were obtained
by optimizing CF states for successively lower layer separations, until
$S$ ceased to increase.

\begin{table}
\begin{center}
\begin{tabular}{c|c|c|c|c|c||c}
$N/2$ &  4  &  5  &  6  &  7 & 8 & $\infty$ \\
\hline
~$S_\text{max}~$ &  ~$0.64$~  &  ~$0.42$~  &  ~$0.38$~  &  ~$0.30$~ &  ~$0.20$~ &  $< 0$ \\
\hline
$d^\text{\tiny CB}$  &   ---   &   $0.87(3)$   &   $0.99(4)$   &  $1.13(5)$    &  --- &   $1.76(11)$   \\
\hline\hline
\end{tabular}
\end{center} \caption{ \label{tab:scaling}
Scaling with system size of the maximal value of the non-zero excitonic superfluid order parameter
for paired CF states $S_\text{max}$ in the absence of CBs.
Pure paired CF trial states are relevant above the layer separation
$d^\text{\tiny CB}$, as discussed in section \ref{sec:MixedCFCBTheory}.
The last column indicates an (overly naive) linear extrapolation over $N^{-1}$
to the thermodynamic limit.
}
\end{table}

This data, together with the values of $d^\text{\tiny CB}$ discussed in
section \ref{sec:pairedResults}, sheds some light on
the question of whether the paired CF state still has the
symmetry of the 111-state in the thermodynamic limit. Given that the
maximal value of the 111 order parameter decreases quickly with $N$ as
summarized in Table \ref{tab:scaling}, it seems that a non-zero
$S$ for paired CF states is a vestige of
finite size systems. Roughly extrapolating $d^\text{\tiny CB}$ in the same
manner confirms this assumption, as it yields a value in the neighborhood
of the onset of the excitonic superfluid order-parameter.
Presumably, the order parameter should thus vanish in the thermodynamic
limit for any state not involving composite bosons. On a more
abstract level, one may reason that interlayer coherence is required
for this order parameter to be non-zero. It seems unlikely that in the thermodynamic limit interlayer CF
pairing alone would achieve this.

With these caveats, our theory supports a
second-order transition between the excitonic superfluid (111 phase) and the paired CF
state, as can be argued from the smooth variation of the order
parameter. Furthermore, for all system sizes that we examined, we
find approximately the same behavior of $S(d)$, which
approaches zero at approximately $d\approx 1.5 \ell_0$. Again, we
interpret the smooth tail of $S(d)$ found above this value
of the layer separation as finite size effects and presume that the
order parameter should approach zero at a precise value $d_c$ in the
thermodynamic limit.

In a recent DMRG-based numerical study,\cite{Yoshioka06} it was
shown that the character of the low-lying excited states changes at
around $d=1.2\ell_0$ for a finite system with $N=24$.    
In light of our results, this transition might correspond to the
layer separation which separates states where
CBs do or do not play a role.  Note that
the value predicted from extrapolation of our results is $d^\text{\tiny CB}(N=24)\approx 1.3\ell_0$.

\subsubsection{CF Pairing Order}
\label{sec:cfpairord}

Assuming that the paired CF phase is distinct
from the excitonic superfluid phase according to the above
hypotheses, there should be a second order parameter that is
particular to the paired CF phase. In analogy with BCS theory, one
would expect an order parameter of the form $\langle
\Psi_\uparrow(r)\Psi_\downarrow(r) \rangle$. However, here we
consider pairing of composite fermions. The important difference is
the Jastrow factors attached to the electrons contribute additional
phase factors. Consequently, a guess for the order parameter
proceeds by unwrapping these phases to give 
\beq 
\exp^{-i \arg [\prod_k (z-z_k)^2]} \exp^{-i \arg [\prod_k (w-w_k)^2]} \; \Psi_\uparrow(z)
\Psi_\downarrow(w), 
\eeq 
where $z$ and $w$ encode the position $r$
in the upper and lower layers respectively. However, pairing is in
the $p$-wave channel and the order parameter is expected to have a
phase that forces it to be zero at coinciding points $z=w$. A
non-zero value might be obtained upon examining operators that are
non-diagonal, i.e.\ $z\neq w$. Though, in such cases the order
parameter continues to have a phase that makes numerical
calculations difficult: averaging a vector rotating arbitrarily in
the plane for different configurations gives a vanishing result. One
must guess the proper phase of the order parameter. For example,
 $\exp [i \arg (z-w)]$ would be appropriate for the $p$-wave case. Thus, we
obtain
\begin{align}
\label{eq:OPPairingProposition}\wp(z,w)=  \exp^{-i \arg [\prod_k(z-z_k)^2]}
\exp^{-i \arg [\prod_k (w-w_k)^2]} \nonumber\\
\times \exp^{-i \arg(z-w)} \Psi_\uparrow(z) \Psi_\downarrow(w).
\end{align} However,
(\ref{eq:OPPairingProposition}) still needs to be modified as numerics require an order
parameter that conserves the particle number in each layer. In principle, one can
multiply (\ref{eq:OPPairingProposition}) by its hermitian conjugate
invoking different positions $\wp^\dagger(z',w')$ to obtain a
candidate for an order parameter satisfying this requirement\beq
\label{eq:NumericOPPairingProposition}
\mathscr{P}=\wp(z,w)\wp^\dagger(z',w'). \eeq This is a rather
complicated operator since it is a function of the four positions $z$, $w$, $z'$
and $w'$. On the sphere, an
additional difficulty arises as a magnetic monopole charge in the
center of the sphere implies the presence of a Dirac string, i.e.\ a
singular point where a
flux tube penetrates the surface of the sphere in order to achieve
magnetic flux conservation. This results in Aharonov-Bohm
phases for wrapping around this point, which must be taken into
account to define $\mathscr{P}$ properly.

We have not yet succeeded to show that a suitably modified BCS order
parameter has a non-zero expectation value for the paired CF states.  However, given the nature of our construction of the wavefunction based on BCS theory, it seems likely that such an order parameter exists.  We hope that in future work we will be able to demonstrate its existence explicitly.

\subsubsection{Pairing Topology}
\label{sec:pairingtopology}
The distinction between the excitonic superfluid phase and the paired CF 
phase should become very obvious on the torus (or periodic boundary condition) 
geometry where the chiral $p$-wave paired phase has a 4-fold topological ground
state degeneracy\cite{ReadGreen,KimNayak01} whereas the 111 phase
has a unique ground state, at least in the thermodynamic limit.   
One would expect that as $d$ is decreased through the phase transition, the 
four-fold degeneracy should split, leaving a unique ground state at small $d$.

\begin{figure}[tttb]
 \begin{center}
\hspace*{-15pt} \includegraphics[width=1.05\columnwidth]{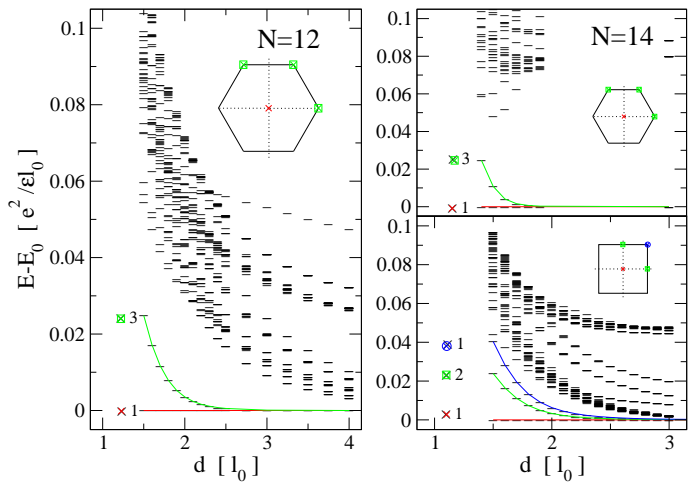}
\end{center}
\caption{  \label{fig:torus}  We display the spectrum of the Coulomb Hamiltonian
on the bilayer torus for different system sizes and geometries as a function of
the layer separation $d$. The lefthand view shows $N=6+6$ particles on a torus
with hexagonal unit-cell, and the righthand view shows $N=7+7$ particles in the
hexagonal (top) and square (bottom) unit cells. The location in the Brillouin
zone of the lowest four energy eigenstates is shown in the respective insets.
The data is suggestive of an approximate fourfold degeneracy of these four
lowest-lying states for intermediate $d$. Strong finite size effects are concluded
from the marked difference of the spectra in the hexagonal and square unit
cells for $N=14$ particles, which precludes strong conclusions about the
thermodynamic limit.
}
\end{figure}

In figure \ref{fig:torus} we show several energy spectra of exact 
diagonalizations\cite{HaldaneRezayiTorus} on the bilayer torus for different 
shaped unit cells and different (even) number of electrons.   This data certainly 
suggests that the lowest four states are separated from the higher energy states 
by a clear gap, and at large enough $d$, these states become degenerate.  
Although suggestive, these data should be viewed with some caution.   What one would 
like to see numerically is that at any $d$ larger than a critical value, the four lowest 
energy states should become increasingly degenerate as the system becomes larger.  
However, this convergence (if present) is not easily seen numerically because of 
discrete shell filling effects.  For example, in the case of the hexagonal lattice 
for $N=14$, at $d=\infty$ the Fermi liquid state is already four-fold degenerate.  
Thus, for this system size and geometry, observation of a four-fold degeneracy should 
not be taken necessarily as evidence of pairing.  Nonetheless, this data is suggestive 
that a phase exists with the topological order that is characteristic of pairing,
i.e., having a fourfold groundstate degeneracy.

\section{Discussion}
\label{sec:discussion}

Perhaps the most crucial question to be answered is the phase
diagram at zero temperature with respect to variations of the
layer spacing $d$. We know for certain that the 111 state is the
ground state at very small $d$ and that two noninteracting composite
fermion Fermi liquids are the ground state for infinite $d$.  We
believe our work sheds substantial light on the intermediate values
of $d$.

 Our work (and also that previously presented in Ref.~\onlinecite{USPRL}) supports the notion that at large but finite values
of $d$ the system is in a $(p_x + i p_y)$-wave paired state of
composite fermions. It has been suggested in
Refs.~\onlinecite{BonesteelPRL96} and \onlinecite{MorinariPRB99}
that even for infinitely weak coupling between the layers there
should be an instability to a paired phase. From our numerical work
it is certainly not possible to determine if the transition to a
paired phase occurs at finite or infinite $d$.  However, it appears
true in our work that the paired trial wavefunction is a notably
better ground state than the unpaired wavefunction even at
reasonably large values of $d \gtrsim 2$.  Since this appears true
even when the inter-layer interaction is weak, and since this phase
appears adiabatically connected to the Fermi liquid,  we should
conclude that this is a weak-pairing phase, rather than a strong
pairing phase.\cite{ReadGreen}   This conclusion is supported by the
fact that, at least at $d > 1$, the occupancy of the orbitals with
small (angular) momentum  (i.e., the inner shells) is higher than
the occupancy of orbitals with higher momentum (the outer shells)
--- this behavior is characteristic of a weak-pairing
phase.\cite{ReadGreen}  Finally, the conclusion of a weak pairing
phase is supported by the topological degeneracies observed on the
torus discussed in section \ref{sec:pairingtopology} above.

At smaller distances between the layers, as discussed above, we
found clear evidence of the order parameter (a broken $U(1)$
symmetry) associated with the 111, or excitonic superfluid phase. We
analyzed this order parameter and found that it approaches zero
smoothly at values close to $d=1.5\ell_0$ with a tail at larger $d$
attributed to finite size effects.  This smooth behavior suggests a
second order transition into the excitonic superfluid phase.
Interestingly, we found that the order parameter can be nonzero even
for our paired $p_x + i p_y$ CF-wavefunctions (with no additional
CBs added to the wavefunction). Our current belief is that this is a
finite size effect, and in the thermodynamic limit, this order
parameter would become nonzero only when the wavefunction has a
nonzero density of composite bosons.  At small layer spacings where
there is a finite value to the excitonic superfluid order parameter,
we find that our wavefunctions with mixed CB-CF and with pairing of the
CFs provide exceedingly good trial states.   It is an interesting
question, which we have not been able to fully answer, whether there
is a distinct (pairing) order parameter associated with the
CF-pairing in the presence of the condensed CBs.

 There are a number of further issues which may be crucially
relevant to experiment which we have not yet mentioned at all and we
will now address briefly.

{\it Finite Temperature and Low Energy Excitations:} Our trial
wavefunction approach is not particularly well suited to studies at
finite temperature. Nonetheless, one could attempt to find trial
wavefunctions for the low lying excited states which would then be
thermally occupied at low but finite $T$.   Certainly, the excitonic
superfluid (111) phase as well as the $p_x + i p_y$  paired CF phase
would have low energy Goldstone modes associated with superfluid
counterflow  (this is essentially a necessary result of having
quantized Hall drag).   Other excitations of these phases should be
gapped, and would be less important at low $T$.   At some higher
characteristic temperature, the order parameters would be destroyed
altogether. It is very possible, that the characteristic temperature
for the paired CF phase would be extremely low, particularly when
the spacing between the layers is large.
Like a superconductor, above this temperature, the putative paired
CF system would behave like a CF-Fermi liquid with some additional
(weak) correlations between the layers.  Of course since this is a
two dimensional system, vortex unbinding physics will be important
and strictly speaking there is no long range order above zero
temperature, and the transition from super to normal would be
smeared to a crossover.

The picture of a mixed CF-CB fluid at small layer spacing discussed
in this paper adds a number of possibilities to the finite $T$ phase
diagram.  For example, one might imagine having a mixture of CFs and
CBs where one or the other species is condensed (but not both).  The
case where the CFs are not condensed, but the CBs are condensed
corresponds with the picture from Ref.~\onlinecite{PRLSimon03} of a
mixed CF-CB fluid where the CFs fill a Fermi sea, but do not pair
(See Appendix \ref{sec:AppFilledShells}). Such a phase could have
low energy excitations associated with excitations of the fermions
around the Fermi surface. We note however that the phase remains
incompressible with respect to ``symmetric" density perturbations
that change the total local charge in both layers.\cite{Milica}  To
understand this incompressibility we simply note that when the total
density compresses, the bosons would then feel an effective
(Chern-Simons) magnetic field (See Eq. \ref{eq:Beff2}), which they
can only accommodate by forming vortices
--- a gapped excitation.  Another way to realize this is to note
that motion of density the entire system (both layers) remains
subject to Kohn's theorem and must only have an excitation at the
cyclotron mode in the long wavelength limit.

Conversely, if one considers a density gain in one layer and a
compensating density loss in the opposite layer, the bosons would
feel no net field.   Although such a density change would presumably
pay the price of the capacitive energy between the two layers, at
long wavelengths such a mode may still be low energy. Indeed, the
superfluid Goldstone mode is of this form.

One might further ask whether there might be any novel low energy
modes in the mixed CF-CB phase associated with motion of CFs in one
direction and CBs in the opposite direction so as to preserve
overall uniformity of charge.   For example, we may  consider the case where
a current of CFs occurs in the same direction in both layers,
such that  $\rho_{\tF}^\uparrow = \rho_{\tF}^\downarrow$ and
$\rho_{\tB}^\uparrow = \rho_{\tB}^\downarrow$ and the total
density in each layer $\rho_{\tB}^\uparrow + \rho_{\tF}^\uparrow
=\rho_{\tB}^\downarrow + \rho_{\tF}^\downarrow$ is a constant. In
this case, there is no capacitive energy, and examining
Eqs.~\ref{eq:Beff} and \ref{eq:Beff2} we see that there is no net
field seen by the bosons, and there is no net field seen by the
fermions. While naively it would appear that such a motion would
yield very low energy modes, it is also possible that the pairing
interaction would couple the motion of the bosons and the fermions,
gapping such a mode even if the fermions are uncondensed.

{\it Layer Imbalance:}  In principle our theory can be generalized
to situations where there are unequal densities in the two layers.
It is well known that the 111 wavefunction can easily accommodate
layer imbalance.\cite{MacdonaldJoglekarPRB02} In the paired-CF
phase, on the other hand, this type of perturbation (like a Zeeman
field in a traditional superconductor) is clearly pair breaking
since the $\uparrow$ and $\downarrow$ Fermi surfaces would be of
different sizes (Although in principle more exotic types of pairing
could be constructed to accommodate such differences). A much more
interesting question to ask is what happens in the regime where
there are both CFs and CBs. The intermediate wavefunctions discussed
in this paper (Eq. \ref{eq:mixedPsi}) do not appear to generalize
obviously to cases where there are unequal numbers of particles in
the two layers (as this would result in a determinant of a
non-square matrix).   We recall that in
Ref.~\onlinecite{PRLSimon03}, mixed CF-CB wavefunctions were
constructed which are identical to those discussed here (with no CF-pairing),
where the antisymmetrization over all particles was done explicitly.
There is no particular difficulty in generalizing that form to cases
with layer imbalance, although such explicit antisymmetrization is
difficult to handle numerically except in very small systems.
Nonetheless, we can at least in principle consider such
generalizations as trial wavefunctions, and we can further consider
allowing pairing of the CFs.  Because of the pairing interaction,
one might guess that the CFs would be stabilized by having equal
numbers of CFs in both layers (as discussed above), and that the
density difference would be accommodated by moving CBs between the
layers.  The fact that experimentally, layer imbalance appears to
stabilize the excitonic superfluid phase,\cite{Spielmanetal04}
suggests further that the transition to this phase coincides with
the appearance of CBs.

{\it Spin:} In the experiments of
Ref.~\onlinecite{EisensteinSpinExperiments} it has been suggested,
that at least in certain samples, the system becomes spin polarized
at low $d$ but is partially polarized at larger $d$.  The transition
is thought to occur near the phase transition to the excitonic
phase. Although all of the trial wavefunctions discussed here have
been for fully polarized systems, they can certainly be generalized
to nontrivial spin configurations. (One should not confuse the
actual spin with the iso-spin, or layer index).   For example, one
could trivially consider having a Fermi sea with some spin down and
some spin up CFs.  Once one considers pairing of this (partially
polarized) Fermi sea, there become many different
possibilities,\cite{ReadGreen} some analogous to superfluid Helium
3. Other exotic possibilities could also occur. For example, one
might imagine two Fermi seas, each pairing in the a $p$-wave
channel. Or one could have unpolarized pairing in an $s$-wave
channel. However, these exotic possibilities may not be
experimentally relevant since the ``superfluid" phase appears to be
polarized,\cite{EisensteinSpinExperiments} suggesting that, as the
spacing between layers is reduced, an unpolarized Fermi sea
condenses into a polarized state (possibly as a first order transition).

{\it Tunneling:}  The wavefunctions we have constructed here are not
only antisymmetric between electrons within a single layer, but are
also antisymmetric between electrons in opposite layers.   As such
these wavefunctions are not particularly destabilized (or
frustrated) by small amounts of interlayer tunneling that destroys
the layer index as a good quantum number.   One should expect,
however, that tunneling between the two layers is quite suppressed
for the CFs since the CF has to carry its Jastrow factor with it,
thereby requiring relaxation of all of the surrounding particles. In
other words, for a CF to tunnel, the entire correlation-hole complex
needs to tunnel with it.   (In yet another language, the effective
magnetic fields in Eq. \ref{eq:Beff} are changed when a CF moves
from one layer to another).   In constrast, tunneling of CBs is
expected to be quite large, since the CB has an identical
correlation hole in each layer.  Indeed, once the CBs are at finite
density, we have found that there is a nonzero expectation of the
excitonic superfluid (111) order parameter which means essentially
that it is uncertain which layer any CB is actually in and the zero
bias tunneling is resonantly enhanced.
With this consideration, we
might expect that tunneling between the two layers will stabilize
the CBs and destabilize the CFs.   When there are CBs present,
tunneling between the layers will also have the effect of gapping
the Goldstone mode, since a particular phase relation is preferred
between the two layers.

{\it Transport:}
Several marked transport phenomena are observed in the bilayer
systems.\cite{PRL84Spielmanetal00,PRL87Spielmanetal01,
PRL88Kelloggetal02,PRL90Kelloggetal03,Spielmanetal04,PRL03Tutuc04}
As discussed above, resonantly enhanced interlayer tunneling current
is a signature of the excitonic superfluid (or 111) order parameter.
In essence, a nonzero value of this order parameter indicates that
in the ground state, each electron is superposed between two layers
and therefore tunneling occurs very strongly, controlled by the
relative phase between the two layers, analogous to the Josephson
effect.

The other two dramatic transport observations are quantized Hall
drag $\rho^D_{xy} = h/e^2$ and superfluid counterflow (which are
very closely related to each other).  In the interlayer-exciton
superfluid (or 111) phase, both phenomena can be understood by the
presence of composite bosons. One argues that superfluid counterflow
derives from coherent transport of CBs or charge-neutral interlayer
excitons. As these objects have no charge, they also do not couple
to the magnetic field and generate no Hall
voltage.\cite{DLReviewNature}

The above reasoning is based on considerations regarding the CB
condensate.  Although our results show that the ``pure" CB
condensate or 111 state occurs only at layer spacing $d=0$, we expect
the transport features of this phase to remain qualitatively similar
to those of the pure 111 state for any sufficiently small $d$ where
the excitonic order parameter (111) remains
nonzero.\cite{PRLSimon03}

Crucially, we note that the two phenomena of quantized Hall drag and
superfluid counterflow would also be observed in a $p_x + i p_y$
paired CF phase, identical to that of the 111 phase --- although
such a CF superconductor would be lacking the strong interlayer
tunneling as discussed above.  [The fact that such a $p$-wave
superconductor shows quantized Hall drag and superfluid counterflow
is easily derived using the technique of Ref.~\onlinecite{ReadGreen}
(See also Ref.~\onlinecite{KimNayak01}) to handle $(p_x + i
p_y)$-wave superconductivity, along with a Chern-Simons
transformation to account for the fact that we are pairing composite
fermions].

It might be interesting to study the Hall drag at interlayer
separations just above the onset of interlayer tunneling. If
experiments  were to identify an intervening regime, which has
quantized Hall drag, but no resonant tunneling, this would be an
indicator of the $p_x + i p_y$ paired CF phase.  Presumably one
would want to examine this transition in high Zeeman field where no
spin transition would complicate experiments. One should be
cautioned, however, that our analysis of transport is very crude. A
more accurate analysis would necessarily involve understanding the
effects of disorder as well as possible edge mode transport, which
has been completely neglected in this work.

\section{Conclusion}
\label{sec:conclusion}

In conclusion, we have derived a composite particle description for
the ground state wavefunction of the quantum Hall bilayer system at
filling factor $\nu=\half+\half$. This ground state is properly
described by interlayer $p$-wave pairing of composite fermions above a
layer separation $d^\text{\tiny CB}$.  More precisely, this pairing
instability occurs in the positive $p$-wave or $p_x+i p_y$ channel.
Below $d^\text{\tiny CB}$, a mixed fluid phase with coexistence of
composite bosons and composite fermions develops, and CBs
successively replace paired CFs upon diminishing $d$. We should
emphasize that positive $p$-wave pairing is the only pairing channel
that is consistent with such a coexistence.

The precision of the composite particle description has the same
order of magnitude as other important trial states in the literature
of the quantum Hall effect, notably as the Laughlin-state at
$\nu=\frac{1}{3}$. The agreement between the trial states and the
exact ground state was checked using energies, overlaps and correlation 
functions, and was found to be in good agreement.

We analyzed the order parameter of the broken U(1) symmetry of the
excitonic superfluid (the 111-state order parameter), and found it
to approach zero smoothly at values close to $d=1.5\ell_0$ with a
tail attributed to finite size effects. We also found this order parameter
to be non-zero for the pure paired-CF-phase. Though we cannot
exclude the contrary with absolute certainty, we believe that this is
a phenomenon occurring only in finite size systems. From the shape
of the order parameter, we conclude that the phase transition between
the 111-excitonic-superfluid phase and the paired CF phase is of second
order. The precise value of the layer separation where this transition
occurs cannot be inferred from our numerics, since the order parameter
continues to be non-zero at all layer separations in small systems.
The transition from the $p$-wave paired CF phase to
an excitonic superfluid phase might also be roughly identified by the
splitting of a 4 fold degeneracy on the torus, indicative of the paired CF phase --- although our finite size torus data needs to be viewed with some caution.

\acknowledgments

The authors gratefully acknowledge helpful discussions with
B.~Halperin, N.~Bonesteel, P. Lederer and N. R. Cooper.
G.M.~would like to thank N.~Regnault for
kind assistance with the DiagHam libraries.
E.H.R. acknowledges support from DOE grant
DE-FG03-02ER-45981.
G.M.~acknowledges support by EPSRC Grant No.~GR/S61263/01.

\appendix

\section{Paired CF wavefunctions on the sphere}
\label{sec:CFCBOnSphere}

The geometry chosen for our numerical calculations is the sphere,
which has the benefit of avoiding boundary effects for finite-size
systems. For our purposes, the most suitable coordinates are the
spinor coordinates \beq \label{eq:spinorCoordinates}
  u=\cos(\theta/2) e^{-i \phi/2}\text{ and }v=\sin(\theta/2) e^{i \phi/2}.
\eeq
In the following, it is convenient to change notations such as to
write particle coordinates with two indices: the
upper index indicates the pseudospin and designates the layer to which it
belongs, whereas the lower index indicates the particle number.
Thus, $(u_i^\sigma,v_i^\sigma)\equiv \Omega_i^\sigma$ describes the
location of particle $i$ with pseudopsin $\sigma$.
The external magnetic field is represented by a magnetic
monopole of strength $N_\phi$ in the center of the sphere, and it is useful
to work in the Haldane gauge.\cite{PRLHaldane83}
In particular, using the formalism of the
stereographic projection between the plane and the sphere,\cite{FanoOrtolani}
one then obtains wavefunctions on the sphere
which can be expressed entirely in terms of $u$'s and $v$'s and contain
no additional phase factors. Our purposes require the translation of
Jastrow factors to the new spinor coordinates on the sphere.
A coordinate $z$ translates to pseudospin up ($\uparrow$) and a
coordinate $w$ translates to pseudospin down ($\downarrow$), e.g., \beq
\label{eq:translationJastrow} (z_i - w_k) \to  (\Omega_i^\uparrow
-\Omega_k^\downarrow) \sim (u^\uparrow_i v^\downarrow_k -
u^\downarrow_k v^\uparrow_i). \eeq Furthermore, the knowledge of a complete set of eigenstates
$\phi_i$ is required to describe (\ref{eq:AllBilayerPaired}) on the
sphere. These eigenstates are given by the monopole harmonics
\cite{Tamm,WuYang,WuYang2} written as $Y_{q,l,m}$ for a total flux
$N_\phi=2q$, and the angular momentum quantum numbers $l = |q| + n$
and $|m| \leq l$. These orbitals are organized in a shell structure
related to the Landau levels on the plane. The LL-index takes
integer values $n=0,1,2,$ etc. Contrarily to the plane, the
degeneracy $d_n$ of these `Landau levels' is not constant but
increasing with $n$ as \beq \label{eq:degSphere} d_n=2(|q|+n)+1.
\eeq
In the thermodynamic limit, $q\to \infty$, whereas $n$ remains finite,
such that the constant LL-degeneracy of the plane is recovered.

\begin{figure}[tttb]
\begin{center}
\psfrag{R1}[c][c]{$R(\theta,\phi)$}
\psfrag{R2longer}[c][c]{$R'(\theta',\phi')$}
\psfrag{G1}[c][c]{$\gamma'$} \psfrag{G2}[c][c]{$\gamma$}
\psfrag{N}[c][c]{$N(0,0)$} \psfrag{t}[c][c]{$\theta_{12}$}
\includegraphics[width=0.45\columnwidth]{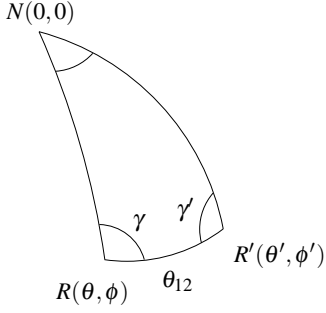}
\end{center}
\caption{  \label{fig:WuYangAngles} Definition of the different
angles for Eq.\ \ref{eq:symmetry_pair_wavefunction} taken from
Ref.~\onlinecite{WuYang2}, adapted to our notations. Points $R$ and $R'$
indicate the positions of two electrons, and a third reference
point can be chosen as the north pole $N$ of the sphere. Generally, the reference
point is given by the singular point of the section for a given
representation of the monopole harmonics.\cite{WuYang}
}
\end{figure}

The proper pair correlation function on the sphere might be deduced
entirely from the requirements of its antisymmetry and the
condition imposed on the flux-count for the resulting bilayer wave
function (\ref{eq:AllBilayerPaired}) to be commensurable with the
111-state. Nonetheless, let us discuss the symmetry of this
two-point function (before projection to the LLL) in more general
terms. A general pair wavefunction on the sphere may be expanded
in terms of monopole harmonics, such that
\begin{align}
  \label{eq:pair_wavefunction_Y_qlm}
&  g\bigl(\Omega_i^\uparrow,\Omega_j^\downarrow \bigr)=g\bigl(\Omega_i^\uparrow-\Omega_j^\downarrow \bigr)=\nonumber\\
&\sum_{n}\sum_m g_{n,m}
Y_{\frac{1}{2},\frac{1}{2}+n,m}(\Omega_i^\uparrow)
Y_{\frac{1}{2},\frac{1}{2}+n,-m}(\Omega_j^\downarrow).
\end{align}
Here, the pair $(\bk,-\bk)$ has been replaced by its analogue on the
sphere $[(n,m),(n,-m)]$. Rotational invariance of
(\ref{eq:pair_wavefunction_Y_qlm}) imposes that $g_{n,m}\equiv g_n$
independent of $m$. In the case of $p$-wave pairing, we must deal
with a slightly more complicated case, since the pair
correlation function is then not rotationally invariant, but rather
acquires a phase. This is reflected by a less restrictive condition
$|g_{n,m}|=g_n$. The angular behavior of
(\ref{eq:pair_wavefunction_Y_qlm}) may then be analyzed according to
Eq.\ 25 from Ref.~\onlinecite{WuYang2}. This equation expresses the sum over the
angular momentum quantum number $m$ of a product of two monopole
harmonics in terms of an amplitude depending solely upon their
distance on the sphere, and a phase depending on several angles. For
our purposes, we need to set $q=q'$, and then take
into account the relationship for the complex conjugation of the
monopole harmonics, (Eq.\ 1 in Ref.~\onlinecite{WuYang2}) in order
to deduce the relationship
\begin{align}
  \label{eq:symmetry_pair_wavefunction}
&  \sum_m (-1)^{q+m} Y_{q,l,m}(\theta',\phi') Y_{q,l,-m}(\theta,\phi)\nonumber\\
  =\,&\sqrt{\frac{2l+1}{4\pi}}\, Y_{q,l,q}(\theta_{12},0)\,
  e^{iq(\phi+\phi')}e^{-iq(\gamma-\gamma'+\pi)}.
\end{align}
This equation holds independently for each shell $n$. The angles $\phi,
\phi', \gamma$ and $\gamma'$ occurring in this expression are named
according to our own conventions and indicated in Fig.~\ref{fig:WuYangAngles}.
The third point of this triangle is a
reference point, that is given by the singular point of the section
on which the monopole harmonics are defined. The phase
$\delta\varphi$, accumulated when taking the two particles around
each other with a small angular separation, may be deduced from the
last term in (\ref{eq:symmetry_pair_wavefunction}). For a half
rotation (i.e. changing the position of both particles), both
$\gamma$ and $\gamma'$ vary by $\pi$, but with different signs,
whereas $\phi$ and $\phi'$ merely change roles. We then have
$\delta\varphi=2\pi q$. Thus, pair wavefunctions expanded in
monopole harmonics $Y_{q,l,m}$ correspond to $2q$-wave pairing,
following the analogy with (\ref{eq:lWavePairing}). The choice of
$q=\half$ for the mixed fluid bilayer wavefunctions is consistent
with the phase of the pair wavefunction found in the 111-state.
Analogously, this may also be concluded from the flux-count argument
introduced at the end of section \ref{sec:MixedCFCBTheory}: naturally,
an orbital $Y_{q,l,m}$ adds a number $N_\phi=2q$ flux to this count.
Thus, with $q=\half$, we recover the previous result that a mixed CF-CB fluid
requires positive $p$-wave pairing of composite fermions.

To summarize, we have outlined how to write the mixed fluid wave
function with paired CF on the sphere. Taking into account the above
considerations, the explicit expression upon adding the projection to
the LLL\cite{PRBJainKamilla97} is:
\begin{align}
  \label{eq:final_psi_trial}
&  \Psi^{\mathrm{CF-CB}}\bigl(\{g_n\}\bigr)
= \det \left[ \frac{J^{zw}_i J^{wz}_j}{u_i^\uparrow v_j^\downarrow-u_j^\downarrow v_i^\uparrow} \,+ \right. \\
&\left. J^{zz}_i J^{ww}_j\sum_{n,m} (-1)^{q+m} g_n \tilde
Y_{\frac{1}{2},\frac{1}{2}+n,m}(\Omega_i^\uparrow) \tilde
Y_{\frac{1}{2},\frac{1}{2}+n,-m}(\Omega_j^\downarrow)
\right].\nonumber
\end{align}
As a reminder, arguments $(\Omega_i^\sigma)$ denote the coordinates
particle $i$ with pseudospin $\sigma$. Jastrow factors must be
expressed following the replacement rule
(\ref{eq:translationJastrow}).

\section{Numerical results for mixed CF-CB states with filled CF shells}
\label{sec:AppFilledShells}

The analysis of the mixed fluid bilayer states with CF pairing
presented in section \ref{sec:numericalResults} has shown that, in
general, the ground state features non-trivial CF pairing. However,
the precise shape of the pairing potential must be found by
optimization over a small set of variational parameters. Since this
requires a considerable numerical effort, it is interesting to
analyze a particular subclass of the mixed fluid states:  those
with filled CF shells. Using the term `shells', we refer to the
spherical geometry, as discussed in Appendix
\ref{sec:CFCBOnSphere}. These filled shell states are obtained
following the choice of parameters (\ref{eq:CFCB_FilledShells}) for
the $g_n$, i.e. choosing very large coefficients up to a reduced
Fermi momentum $(k_F)_\tF$ to force the respective number of
electrons into CF orbitals. Remaining electrons then occupy CB
states.

Given the degeneracy of CF shells on the sphere
(\ref{eq:degSphere}), with $q=\half$ for the mixed fluid states,
there are a small number of possible filled shell states for each
system size $N$. Explicitly, the series of possible CF numbers per
layer for $n_s$ filled CF shells is given by
\begin{equation}
  \label{eq:filled_shell_sequence}
  N_{1\text{\tiny F}}(n_s)=n_s(n_s+1)= 2,6,12,20,\ldots\, .
\end{equation}
Though these filled shell states are known not to be ground states
of the bilayer system, they represent intermediate states between
the 111-state and the CFL, and are better approximations of the
ground state than either of the latter two states for intermediate
layer separations.

As an example, $\state{2}$ as described in Ref.~\onlinecite{PRLSimon03}
is such a filled shell state without CF pairing. In order to show that our calculation
reproduces exactly the state $\state{2}$ for large $g_0$, and
$g_n=0,\forall\, n\geq 1$, we have calculated the overlap of 
that state with a special case of our trial states (with $g_0$ large and all other $g_n=0$),
and find it to be precisely equal to one within the numerical precision
of our calculation: $|\langle 2\text{ Fermions} | \Psi^\text{CF-CB}(g_0\to\infty)\rangle|^2=
0.9999999 \pm 10^{-6}$, for an overlap integral evaluated with $5\times 10^5$ Monte-Carlo samples.

The agreement we have found
supports our claim that we can indeed generate precisely the mixed CF-CB states introduced in Ref.~\onlinecite{PRLSimon03}
using our single determinant wavefunctions.  This agreement further supports our interpretation of the $g_n$
as controlling the occupation probability of the respective CF shell.   Note that when
choosing $g_n$ to be large, this means that the respective CF states
are inert (i.e., the orbitals are fully filled and they do not participate in nontrivial pairing).
It then does not matter whether the pair correlation function is chosen symmetric or antisymmetric.

\begin{table*}[ttbp]
\begin{center}
\begin{tabular}{rcccccccccccccccccccc}
\hline\hline
$N_1$ & ~~~ & \multicolumn{2}{c}{1 filled shell} & ~~~ & \multicolumn{5}{c}{2 filled shells} & ~~~ &\multicolumn{8}{c}{3 filled shells} & ~~~ & $\delta \left[\sum p \right] $ \\
\hline
 && $p(0)$ & $\delta p_0$ &  & $p(0)$ & $\delta p_0$ && $p(1)$ & $\delta p_1$ &  & p(0) & $\delta p_0$ && $p(1)$ & $\delta p_1$ && $p(2)$ & $\delta p_2$ & &  \\
\hline
5 & & 0.400974 & +0.24 &  & 0.400877 & +0.22 && 0.604851 & +0.81 & &  \multicolumn{8}{c}{N/A} & & +0.57 \\
\hline
6 & & 0.334152 & +0.25 &  & 0.334157 & +0.25 && 0.678183 & +1.73 & &  \multicolumn{8}{c}{N/A} & & +1.23 \\
\hline
7 & & 0.286419 & +0.25 &  & 0.286421 & +0.25 && 0.581292 & +1.73 &  & 0.285971 & +0.09 && 0.575032 & +0.63 && 0.142858 & $\pm$0.0 & & +0.94 \\
\hline
8 & & 0.250619 & +0.25 &  & 0.250620 & +0.25 && 0.508638 & +1.73 &  & 0.250559 & +0.22 && 0.500022 & $\pm$0.0 &&  0.258746 & +3.5 &  &  +0.93 \\
\hline
12 & & 0.167082 & +0.25 &  & 0.167082 & +0.25 && 0.339134 & +1.74 &  &  0.167082 & +0.25 && 0.339158 & +1.75 && 0.522862 & +4.57 & &  +2.91 \\
\hline\hline
\end{tabular}
\caption{Occupations $p(n)$ calculated according to Eq.
\ref{eq:occupationProba} for filled shell states. At a given system
size $N_1=N/2$, values for sample calculations of all possible
filled shell states are indicated. The deviations of $p(n)$ from the
expected occupation probabilities $\delta p_n=[ p(n) /
(N_{1\tF}(n)/N_1) ]- 1$  are indicated in percent. The last column
gives the percent deviation from $1$ for the sum of occupation
probabilities of the state with the maximal number of filled shells.
} \label{tab:FilledOccupations}
\end{center}
\end{table*}

In the case of filled CF shells, one can argue that our paired CF
description and the mixed fluid picture from Ref.~\onlinecite{PRLSimon03}
are identical. However, we also find perfect agreement for the 
state where all electrons occupy CF orbitals,
$\state{5}$, which is not a filled shell configuration: the overlap
of the corresponding trial state with the explicitly constructed 
CFL state $\state{5}$ was
found to be $|\langle 5\text{ Fermions} | \Psi^\text{CF-CB}(g_0,g_1\to\infty)\rangle|^2= 
0.999991 \pm 3\times 10^{-5}$ (evaluated over $10^6$ Monte-Carlo samples).
As opposed to the previous case, in order to obtain this
agreement, it is required that the pair correlation function $g_\tF$
be chosen antisymmetric (see Appendix \ref{sec:CFCBOnSphere}). As pointed
out in the main text, and discussed previously in Ref.~\onlinecite{USPRL}, 
this agreement is possible only for cases where the CF-sea deviates
from a filled shell configuration by at most one electron per layer.

Since the fraction of CFs and CBs is known for the mixed fluid
states, these represent a testing ground for the validity of
Eq.\ \ref{eq:occupationProba}. Numerical evaluation indeed confirms that
the correct fraction of CFs $p(n_s)=N_\tF(n_s)/N_1$  is obtained from
(\ref{eq:occupationProba}) within about one percent error
(See Table \ref{tab:FilledOccupations}).
Typically, when calculating a Fermi liquid state, $\sum_{n_s}p(n_s)$
is slightly larger than one but remains within the same error
margin.

Having clarified that the filled CF shell states represent a
subclass of the mixed fluid states in Ref.~\onlinecite{PRLSimon03}
but with the advantage that the representation (\ref{eq:final_psi_trial}) is
computationally easier to evaluate, we may study this class of
states up to very large system sizes.

We have studied larger systems, focusing our attention to system sizes of
the sequence (\ref{eq:filled_shell_sequence}). For a system size
corresponding to $n_s$ filled shells, we may construct $n_s+1$
different trial states, notably the 111-state and the states with
$1,2,\ldots,n_s$ filled shells. The state with all shells filled
(i.e. the CF Fermi liquid) gives us a criterion to test whether the
parameters $g_n$ have been chosen large enough to transform all
particles to composite fermions. Such a state features
no interlayer correlations and, consequently, its interlayer
correlation function should be constant. All one needs to do is
to tune the $g_n$ until this situation is reached. Empirically,
we have found that values $g_n\gtrsim 1000$ satisfy this criterion.

The biggest system analyzed in this way had $N=42+42$ particles.
As the the exact ground state energy is not known for such large
systems, we only compared the different filled shell states.
At zero layer separation, the 111 state is the exact ground state.
Interestingly, states with a small number of CFs have a very large
overlap with the 111-state, such that MC simulations have difficulty
in resolving their difference in energy. However, there is a general
tendency that states including CFs have lower energy at increasing $d$.
This suggests that a finite fraction of CF could eventually be favorable at
any finite $d$ in the thermodynamic limit. Going from small to larger layer
separations, states with subsequently more filled CF-LLs clearly become
the most favorable trial states.

\begin{figure}[ttb]
  \begin{center}
    \includegraphics[width=.8\columnwidth]{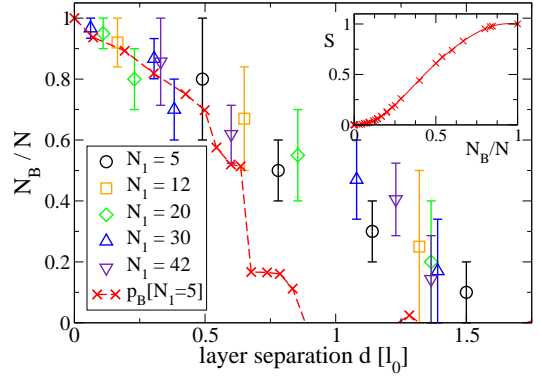}
  \end{center}
\caption{  \label{fig:ratio_of_bosons}  Intersections of the energy
levels of filled CF shell states allow deduction of estimates for
the range of the most favorable fraction of CB at the layer
separation $d^\times$ of the point of intersection. The results for
various system sizes as small as $N_1=5$ and as large as $N_1=42$ electrons
are represented collectively in this plot, and show a good
coherence. Note that CF are formed in the system at very small $d$.
A rough linear extrapolation of these results suggests
that the ratio $N_\tB/N$ would vanish at approximately
$d=1.7\ell_0$. However, CF-pairing changes these figures, as highlighted
by the occupation probability $p_\tB$ for $N_1=5$ that is given for comparison.
The inset shows the monotonic relationship between the
order parameter $S$
and the fraction of bosons for data from various filled shell states
at different $N$.}
\end{figure}

The layer separations $d^\times_{n_s}$, where we observed the
level crossings between a first state with $n_s-1$ filled CF shells and
a second one with $n_s$ shells filled are spread out over a large interval
of layer separations ranging from $d^\times_0 \lesssim 0.05\ell_0$ to
$d^\times_5 \sim 1.5$. As stated before, neither of the filled shell states describes the
ground state of the system at the point of their level crossing.
Nonetheless, the $d^\times_{n_s}$ provide an estimate of the range
of $N_\tF/N$ that would best characterize the ground state at this
layer separation in the absence of CF-pairing. From this kind of reasoning,
we can infer that
\begin{equation}
\frac{N_\tF(n_s-1)}{N} \lesssim \left.\frac{N_\tF}{N}\right|_{d^\times_{n_s}}
\lesssim \frac{N_\tF(n_s)}{N}.
\end{equation}
Collecting data from level crossings $d^\times_{n_s}$ at different system sizes,
we established Fig.\ \ref{fig:ratio_of_bosons}, where we have represented the
complementary ratio of composite bosons $N_\tB/N=1-N_\tF/N$. For the filled shell states,
the ratio $N_\tB/N$ is related to the order parameter $S=|\langle \mathcal{S} \rangle|$
via a monotonically growing function  (See the inset of Fig.\ \ref{fig:ratio_of_bosons}).

Given that CF pairing predominantly lowers the energy of states that
contain a substantial fraction of CFs, the range for $N_\tB/N$
indicated in Fig.\ \ref{fig:ratio_of_bosons} should be seen as an
estimate for the upper bound of the fraction of bosons. This is most
drastically illustrated by the occupation probability of CB orbitals
$p_\tB$ (for $N=5+5$ particles) that is given for reference in this figure.
At small layer separations, where the mixed fluid description is at work, this
curve is within the error bars deduced from the filled shell
analysis. However, once the paired regime is approached, the true
occupation of boson orbitals drops rapidly and the estimate made
here clearly overestimates the actual value.

\section{Numerical Methods}
\label{sec:numericalmethods}

As stated in section \ref{sec:numericalResults}, the aim of our
numerical simulations of the bilayer states (\ref{eq:mixedPsi}) was
to show that they potentially represent the ground state. However,
to achieve an explicit representation of the ground state at a given
layer separation $d$, one must find the corresponding set of
variational parameters $\{g_n,c_\tB\}_d\equiv\gamma_d$ that yields
an optimal trial state (assuming a time reversal invariant
interaction, all LLL wavefunctions can be written as polynomials
with real coefficients, so $\{g_n\}$ were considered real). 
This was realized either by maximizing the overlap
with the exact ground state, or by minimizing the energy. 
Both operations represent non-trivial optimization problems. 

In general, optimization algorithms require a large number of function
evaluations before obtaining a good `guess' of the optimal solution.
Furthermore, our calculations were based upon Monte-Carlo simulations, a
statistical method which yields statistical errors vanishing only as
the inverse square root of the number of samples.
This means that any optimization method is bound to make a
trade-off between the uncertainty it allows for the precision of
function evaluations and the number of such evaluations it requires.

Monte-Carlo sampling to evaluate expectation values is used in both
methods below. Na\"ively, each set $\trialIdx=\{g_n,c_\tB\}$ requires a separate
Monte-Carlo simulation, though it is possible by using correlated
sampling to simulate at the same time many choices of these parameters.
A prerequisite for correlated sampling is that correlations in the 
simulated wavefunctions are similar, such that the ensemble of 
samples used is similarly relevant to all of them.
With this approach, it is easy, for example, to numerically evaluate local derivatives 
with respect to the variational parameters. Best
results for our calculations were achieved by using a self-consistent
sampling function $F$ --- an expression obtained as a Jastrow product
form exploiting the correlation functions\cite{NoteCorrelations}
$\corrSym_{\sigma\sigma'}^\trialIdx$ calculated in the same run (let
the superscript $\trialIdx$ be a reference of a distinct trial
state). This yields \beq
  F = \prod_{i<j} \corrSym_\uu^\trialIdx(z_i-z_j) \prod_{i<j} \corrSym_\dd^\trialIdx(w_i-w_j) \prod_{i,j}
 \corrSym_\ud^\trialIdx(z_i-w_j),
\eeq where $\corrSym_\uu(r)=\corrSym_\dd(r)$ by symmetry. The most 
important part in this Ansatz are the interlayer correlations 
$\corrSym_\ud$, as the intralayer correlations are rather similar 
for \emph{all} possible choices of the parameters $\trialIdx$.

\subsection{Energy optimization}

Due to the statistical errors that underly the Monte-Carlo
simulations, computation time increases as the inverse square
of the required precision, such that any optimization scheme using local
derivatives of the energy is difficult. Iterative comparison of
neighboring states in correlated sampling is a slow route to
optimization.
As the results shown in section \ref{sec:numericalResults} unveil,
the difficulty of finding a good optimization scheme suitable 
for our case is also due to the inherently good correlations common 
to all trial functions: further improvement only concerns rather 
small relative differences in energy.

To meet these challenges, we successfully deployed a rather subtle optimization 
method\cite{UmrigarLinear} based on iterated diagonalization of the
Hamiltonian in the space spanned by the present trial state
$\ket{\Psi_0}$ and its derivatives with respect to the variational
parameters $\ket{\Psi_i}=\derivX{\gamma_i}\ket{\Psi_0}$. The
trial-state representation for the next iteration can be
represented as the Taylor expansion \beq
 \ket{\Psi} = \sum_{i=0}^{n_c} c_i \ket{\Psi_i},
\eeq where $c_i$ is the proposed change in the
parameters. The values $c_i$ may be obtained as the solution of
the generalized eigenvalue problem in this non-orthogonal and
incomplete basis \beq
 H \ket{\Psi} = ESc,
\eeq where $S$ is the overlap matrix $S_{ij}=\langle
\Psi_i\ket{\Psi_j}$. Even better results were obtained using a
slightly different basis which was additionally chosen to be
semi-orthogonalized with respect to $\ket{\Psi_0}$, such that
$\langle \Psi_0\ket{\Psi_i}=0, i=1,\ldots,n_c$. The stabilisation
of this procedure is discussed in Ref.~\onlinecite{UmrigarLinear}.

\subsection{Optimization of Overlaps}

Where the exact ground state is known from exact diagonalization,
we may revert to a simpler method of singling out the optimal
trial wavefunction of the form~(\ref{eq:mixedPsi}), namely optimizing 
the overlap with the exact wavefunction. Here, the main difficulty
lies in the evaluation of the overlap between the trial states
and the exact ground state: the trial wavefunctions are known
in real-space, results from exact diagonalization are given in the
second quantized notations of occupation in momentum space
\beq
\label{eq:ExactDiagState}
|\Psi_\text{exact}\rangle = \sum_{\alpha=1}^{D(L_z=0)} a_\alpha c^\dagger_{m_1(\alpha)\uparrow}\cdots
c^\dagger_{m_{N_1}(\alpha)\uparrow} c^\dagger_{m_1(\alpha)\downarrow}\cdots
c^\dagger_{m_{N_1}(\alpha)\downarrow}|0\rangle,
\eeq
with $\alpha$ denoting a many-body basis state in the Hilbert subspace at
$L_z=0$ of dimension  $D(L_z=0) \gg D(L=0)$, and corresponding
amplitudes $a_\alpha$. Converting trial states $\Psi_\text{trial}(\{z_i,w_i\})$
into the second quantized basis is difficult, so overlaps are 
calculated in real space with a Monte-Carlo evaluation of the integral over
many configurations $\sigma = (\{z_i,w_i\})$
\beq
\langle \Psi_\text{exact}|\Psi_\text{trial}\rangle=
\int\,d\sigma \, \Psi^*_\text{exact}(\sigma)\,\Psi^\trialIdx_\text{trial}(\sigma)
\eeq
based on 
$\Psi_\text{exact}(\{z_i,w_i\}) = \langle \{z_i,w_i\} |\Psi_\text{exact}\rangle$.
Each evaluation of $\Psi_\text{exact}(\sigma)$ requires the evaluation of
$D(L_z=0)$ Slater determinants of size $N_1$, which is the most time-consuming
operation. It is therefore advised to generate a sequence of Monte-Carlo
samples from the exact wavefunction only \emph{once}, and subsequently use it to 
calculate overlaps with various trial wavefunctions via correlated sampling.

The optimization step of finding the value for the parameters $\trialIdx_\text{opt}$
which yields the highest overlap turns out to be rather simple. The Fletcher-Reeves
method (a steepest descent algorithm) was found to yield satisfactory results.

\section{Analysis of correlation functions}
\label{sec:correlations}
In the main text of the paper, we analyzed the energies of trial states and their
overlaps with the exact trial state as a measure of their performance. Alternatively,
one may use a comparison of the correlation functions as a gauge for capturing the
physics of the exact solution. The correlation functions provide more information
about the system, which makes them a more comprehensive measure, but also more
difficult to interpret than a single number as the energy.
We define the pair correlation functions\cite{NoteCorrelations} as
\begin{equation}
    \corrSym_{\sigma \sigma'}(\theta) = \frac{{\mathcal N}_{\sigma \sigma'}}
{\langle \rho_\sigma\rangle \langle \rho_{\sigma'} \rangle}
\langle  \rho_\sigma(\vec r) \rho_{\sigma'}(\vec r') \rangle
\end{equation}
where $\rho_\sigma(\vec r)$ is the density in layer $\sigma$ at
position $\vec r$,  and  $\theta$ is defined as the great circle
angle between positions $\vec r$ and $\vec r'$. The normalization is
chosen such that $\corrSym(r\to\infty)={\mathcal N}_{\sigma
\sigma'}$, with \beq {\mathcal N}_{\sigma \sigma'} =
[(N/2)-\delta_{\sigma \sigma'}]/(N/2). \eeq 
This choice accounts for the different number of interacting particle
pairs in the interlayer and intralayer correlations.

\begin{figure}[ttbp]
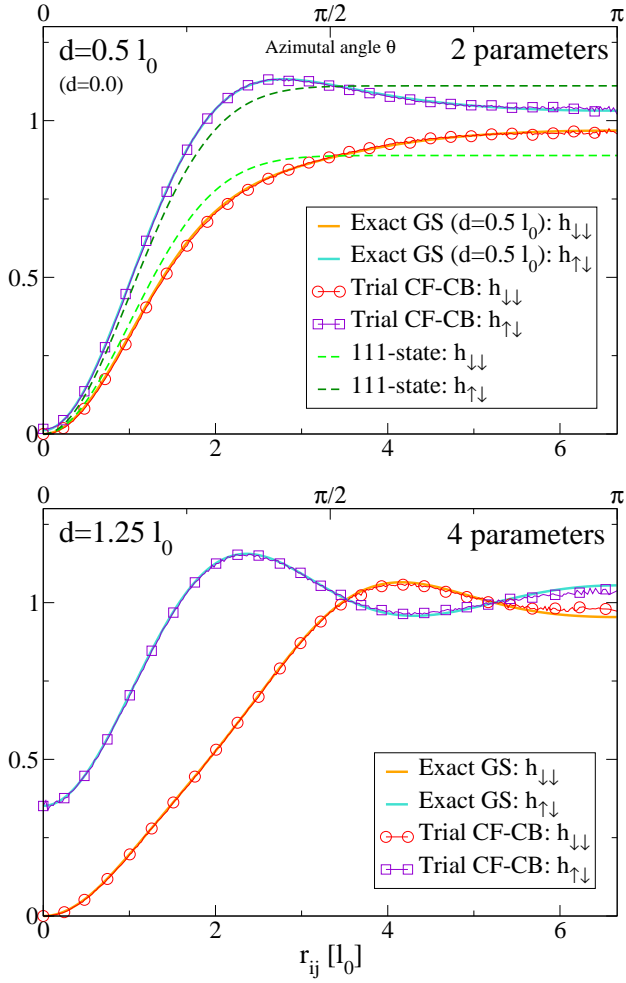

  \begin{center}
    \includegraphics[width=.95\columnwidth]{correlations_N5_d0_0_0_5}\\[9pt]
    \includegraphics[width=.95\columnwidth]{correlations_N5_d1_25} 
\caption{  \label{fig:CorrelationFunctions}  Correlation functions
$\corrSym_\ud(r)$ of the respective ``best'' trial states at
layer separations $d=0,\,0.5\ell_0$ (top) and $d=1.25\ell_0$ (bottom) 
for a system with $5+5$ electrons.
The agreement between the trial wave-functions (thin lines with
symbols) and the exact results (bold lines) is significant for any
value of $d$. Even at small finite $d$, the correlations
$\corrSym_{\sigma\sigma'}(r)$ differ noticeably from those of the
111-state (top). 
The number of variational parameters employed is indicated in each case. 
The small noise in some of the curves for the trial states is Monte-Carlo error.
}
  \end{center}
\end{figure}

Let us first discuss some of the physics revealed by the correlation
functions in the bilayer (see also Ref.~\onlinecite{Yoshioka06}).
Some correlation functions are shown in Fig.\ \ref{fig:CorrelationFunctions} 
for $N=5+5$ electrons. 
The top panel of Fig.~\ref{fig:CorrelationFunctions} shows both
the correlation functions at $d=0$ and $d=0.5\ell_0$.
Note that for $d$ as small as $d=0.5\ell_0$, the correlation hole
for small $r$ in the intralayer correlation function $\corrSym_\dd$
is noticeably enlarged with respect to the 111-state, the exact ground
state at $d=0$. The correlation hole in the interlayer
correlation function $\corrSym_\ud$ is reduced accordingly, with
$\corrSym_\ud(0)>0$.  We find that
the observed change in the correlation functions can be understood
by exclusively considering the admixture of CF to the 111-state:
the mixed fluid wavefunctions (\ref{eq:mixedPsi}) perfectly reproduce
these correlations.

With growing $d$, the anti-correlations described by the correlation
hole in $\corrSym_\ud$ continue to decrease and the correlation hole
in $\corrSym_\dd$ expands to its full size. For choices of $g_n$
that correspond to sufficiently large numerical values such that the
correlation hole in $\corrSym_\dd$ has reached its full size, the
shape of the intralayer correlation function is relatively
insensitive to the precise values of these parameters. This means
that intralayer correlations are coded into the Jastrow factors
regardless of the specific (projected) CF orbital. In contrast, the
interlayer correlation function $\corrSym_\ud$ has a strong
dependence on the shape of $g_n$.

\begin{figure}[tttt]
  \begin{center}
    \includegraphics[width=.9\columnwidth]{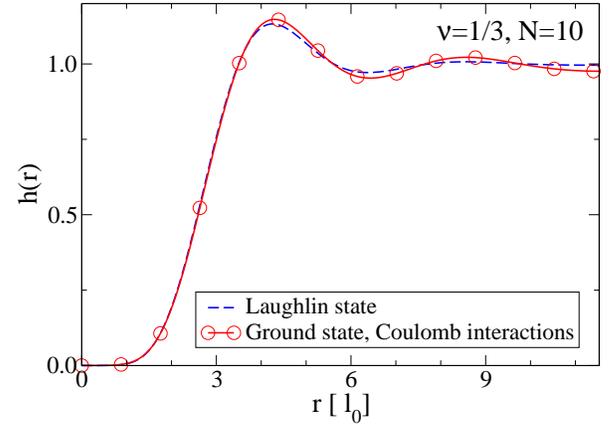}
    \caption{\label{fig:Laughlin} 
    Two-point correlation function $h(r)$ of the Laughlin state as it compares to that of the exact
    ground state for Coulomb interactions on the sphere. The good trial energy
    of the Laughlin state arises from the precise reproduction of the exact
    correlations at short layer separations. However, oscillations at larger layer separation
    have a stronger amplitude in the ground state. }
  \end{center}
\end{figure}

In in section \ref{sec:pairedResults} of the main text, we highlighted 
that paired states without
an admixture of CB correlations reproduce exact ground states down to
$d\sim\ell_0$. As an example for a correlation function $\corrSym_\ud$
in this regime, the bottom panel of Fig.\ \ref{fig:CorrelationFunctions},
showing $d=1.25 \ell_0$, 
features a strong anti-correlation of electrons in both layers. This
correlation hole can thus be explained entirely in terms of
CF-pairing, which seems counterintuitive as one would expect pairing
to enhance correlations between the layers. With regard to the shape
of the pair wavefunctions (\ref{eq:lWavePairing}) where
$g(z,w)\propto (z-w)$ for $p$-wave pairing, we can more clearly
understand this feature. By virtue of this property, $p_x+ip_y$
pairing introduces interlayer \emph{anti}-correlations on short
length scales. As the pair wavefunction is forced to have a maximum
and to decay for $r\to\infty$, $g$ is guaranteed to describe a bound
state of pairs with some finite typical distance between the bound
particles. Correspondingly, the correlation hole in $\corrSym_\ud$
is accompanied with an enhanced correlation around $r\approx
2\ell_0$.

In the regime of intermediate layer separation shown in the bottom 
panel of Fig.\ \ref{fig:CorrelationFunctions}, the overlap
with the exact ground state is not quite perfect [the state shown
was optimized on the overlap, attaining a value of $0.987(3)$ for 
$d=1.25\ell_0$]. Optimization over either the overlap or the energy
results in highly accurate correlation functions at short distances,
while the large $r$ behavior is weighted lower and may not be fully
reproduced. However, as shown in Fig.\ \ref{fig:CorrelationFunctions}, 
the correlations of the paired CF-CB states are extremely close to the
exact correlation functions. For these variational state, some of the 
accuracy at short distances can be traded for a better reproduction
of the large $r$ behavior.

As a prominent reference case, we might cite the correlation function
of the Laughlin state. Though the Laughlin state is a very accurate
description of the ground state at filling factor $\nu=1/3$, its correlation
function still deviates noticeably from the correlation of the exact ground
state at large $r$, as shown in Fig.~\ref{fig:Laughlin}.

\bibliography{double_layer}

\begin{thebibliography}{67}
\bibitem[{Til()}]{TildeNotation}
\bibinfo{note}{In Ref.~\onlinecite{USPRL}, we had used an abbreviated notation
  with $\tilde\phi_\bk$ including Jastrow factors $J_i$. Here, we differently
  chose to write the Jastrow factors explicitly to highlight the different flux
  attachments leading to CFs and CBs.}
\bibitem[{Not({\natexlab{b}})}]{NoteCorrelations}
\bibinfo{note}{We choose to represent the pair correlation function using the
  symbol $h(r)$ for the entirety of this paper, since $g(r)$ is used as a
  symbol for the pair wavefunction.}
\end{thebibliography}

\end{document}